%% file: main.tex
\documentclass[letterpaper,twocolumn,10pt]{article}
\usepackage{usenix2019_v3}

\input{packages}

\input{./acronyms}
\input{./macros}

\begin{document}

\date{}
\pagestyle{empty} 

\title{SoK: What Don’t We Know? Understanding Security Vulnerabilities in SNARKs}


\author{
{\rm Stefanos Chaliasos}\\
Imperial College London
\and
{\rm Jens Ernstberger}\\
\and
{\rm David Theodore}\\
Ethereum Foundation
\and
{\rm David Wong}\\
zkSecurity
\and
{\rm Mohammad Jahanara}\\
Scroll Foundation
\and
{\rm Benjamin Livshits}\\
Imperial College London \& Matter Labs
}

\maketitle              

\begin{abstract}
Zero-knowledge proofs (ZKPs) have evolved from being a theoretical concept providing privacy and verifiability to having practical, real-world implementations, with SNARKs (Succinct Non-Interactive Argument of Knowledge) emerging as one of the most significant innovations. Prior work has mainly focused on designing more efficient SNARK systems and providing security proofs for them. Many think of SNARKs as ``just math,'' implying that what is proven to be correct and secure \emph{is} correct in practice. In contrast, this paper focuses on assessing end-to-end security properties of real-life SNARK \emph{implementations}. 
We start by building foundations with a system model and by establishing threat models and defining adversarial roles for systems that use SNARKs. 
Our study encompasses an extensive analysis of~141 actual vulnerabilities in SNARK implementations, providing a detailed taxonomy to aid developers and security researchers in understanding the security threats in systems employing SNARKs. 
Finally, we evaluate existing defense mechanisms and offer recommendations for enhancing the security of SNARK-based systems, paving the way for more robust and reliable implementations in the future.

\end{abstract}

\section{Introduction}

\ZKPs have undergone a remarkable evolution from their conceptual origins in the realm of complexity theory and cryptography~\cite{goldwasser1985knowledge, goldwasser1989knowledge} to their current role as fundamental components that enable a wide array of practical applications~\cite{ernstberger2024zkps}. Originally conceptualized as an interactive protocol where an untrusted prover could convince a verifier of the correctness of a computation without revealing any other information (zero-knowledge)~\cite{goldwasser1985knowledge}, \ZKPs have, over the past decade, transitioned from theory to practical widely used implementation~\cite{sasson2014zerocash,bowe2020zexe,ozdemir2022experimenting,bonneau2020mina,miers2013zerocoin,pertsev2019tornado,protocolLabs2017filecoin,world2019verifiable}.

On the forefront of the practical application of \textit{general-purpose} \ZKPs are \SNARKs~\cite{gennaro2013quadratic, parno2016pinocchio, groth2016size,gabizon2019plonk,chiesa2020marlin}.
\SNARKs are non-interactive protocols that allow the prover to generate a succinct proof. The proof is efficiently checked by the verifier, while maintaining three crucial properties: completeness, soundness, and zero-knowledge.
What makes \SNARKs particularly appealing is their general-purpose nature, allowing any computational statement represented as a \emph{circuit} to be proven and efficiently verified.
Typically, \SNARKs are used to prove that for a given function $f$ and a public input $x$, the prover knows a (private) witness $w$, such as $f(x,w)=y$. 
This capability allows \SNARKs to be used in various applications, including ensuring data storage integrity~\cite{protocolLabs2017filecoin}, enhancing privacy in digital asset transfers~\cite{miers2013zerocoin,sasson2014zerocash} and program execution~\cite{bowe2020zexe,bonneau2020mina}, as well as scaling blockchain infrastructure~\cite{polygon2023,polygon2023miden,era2023,scroll2023}. Their versatility also extends to non-blockchain uses, such as in secure communication protocols~\cite{zhang2020deco, lauinger2023janus} and in efforts to combat disinformation~\cite{kang2022zkimg, ko2021zkimageredacting}. Unfortunately, developing and deploying systems that use \SNARKs safely is a challenging task.

In this paper, we undertake a comprehensive analysis of publicly disclosed vulnerabilities in \SNARK systems. Despite the existence of multiple security reports affecting such systems, the information tends to be scattered. Additionally, the complexity of \SNARK-based systems and the unique programming model required for writing \ZK circuits make it difficult to obtain a comprehensive understanding of the prevailing vulnerabilities and overall security properties of these systems. 
Traditional taxonomies for software vulnerabilities do not apply in the case of \SNARKs; hence, we provide the seminal work that addresses this gap by providing a holistic taxonomy that highlights pitfalls in developing and using \SNARKs.
Specifically, we analyzed~\empirical{\totalvulns} vulnerability reports spanning nearly~$6$ years, from~2018 until~2024. Our study spans the entire \SNARK stack, encompassing the theoretical foundations, frameworks used for writing and compiling circuits, circuit programs, and system deployments. We systematically categorize and investigate a wide array of vulnerabilities, uncovering multiple insights about the extent and causes of existing vulnerabilities, and potential mitigations.

\vspace{1cm}

\point{Contributions} 
\nopagebreak
\begin{itemize}
\item \textbf{\SNARKs system and threat models:} We provide the first framework for reasoning about systems built using \SNARKs, analyzing interactions between different components, defining adversaries and their knowledge, and discussing potential implementation-level vulnerabilities and their impact.
    
\item \textbf{Study of vulnerabilities:} We present the first systematic study of known vulnerabilities in systems using \SNARKs. We gathered~\totalvulns vulnerabilities from~\totalaudits audit reports, $16$ vulnerability disclosures, as well as a number of bug trackers of popular \SNARK projects. When it comes to \SNARKs, this is the first study of this scale in the literature. Further, because of the breadth of our coverage, we believe our findings to be representative of the entire \SNARK space.
    
\item \textbf{Vulnerabilities taxonomy:} We introduce a taxonomy for classifying vulnerabilities in \SNARKs, highlighting unique vulnerabilities and common pitfalls in the \SNARK stack that help researchers and practitioners better understand important threats in the \SNARK ecosystem. 
    
\item \textbf{Analyzing defenses:} We analyze the main defense techniques proposed by the research and practitioner communities and highlight some notable gaps.
\end{itemize}

\point{Key Findings}
We find that developers seem to struggle in correctly implementing arithmetic circuits that are free of vulnerabilities, especially due to most tools exposing a low-level programming interface that can easily lead to misuse without extensive domain knowledge in cryptography.
In detail, we find the following flaws to be most pressing:
\emph{(i)} \textit{Implementation bugs across \SNARK systems' layers}, including classic vulnerabilities like input validation errors and over/underflows. These can undermine \SNARKs' core properties: completeness, soundness, and zero-knowledge.
\emph{(ii)} \textit{The unique programming model for \SNARK circuits poses challenges, often leading to under-constrained circuits}. This category emerges from overlooking constraints or misinterpreting logic into circuits. The low-level nature of \SNARK \DSLs, such as Circom, exacerbates vulnerabilities due to a lack of common high-level programming features such as basic types.
\emph{(iii)} \textit{Design and implementation errors in proof systems} are critical yet often overlooked vulnerabilities. These errors may originate from the frameworks implementing the proof systems, like an implementation error in Gnark's Plonk verifier, or from the theoretical foundations themselves. An example is the ``Frozen Heart'' vulnerability, attributed to incomplete descriptions in the original proof system papers, leading to significant implementation errors~\cite{dao2023weakfiatshamir}.

It is important to highlight that \ZK systems are \emph{not} ``just math''~---~they are complex, ``compositional'' systems where cross-layer interactions can introduce complex vulnerabilities. This paper attempts to cover the entire gamut of erroneous possibilities in the \ZKP space.


\section{Background on SNARKs}
\label{section:background}

A \ZKP enables an entity to prove that a statement is true, without disclosing anything besides the veracity of the statement.
A \ZKP is termed a \ZK-SNARK if the proof size and verification time are sublinear in the statement to be proven, the communication between prover and verifier is non-interactive, and security holds against a computationally bounded prover.
Common \ZK-\SNARKs are targeting the problem of \textit{circuit satisfiability}, where the statement is represented as an arithmetic circuit. Hence, general-purpose \SNARKs prove a fixed NP relation $\mathfrak{R}$, and allow the prover to convince a verifier that for the public input $x$ they know a witness $w$ such that $(x, w) \in \mathfrak{R}$. 
\emph{Pre-processing \SNARKs}~\cite{chiesa2020marlin}, which are the focus of this work, additionally introduce a  \emph{setup} phase that encodes the relation being proven into a succinct representation. 

In summary, a \SNARK is composed of three algorithms:

    \par\noindent $\textit{Setup}(pp) \rightarrow (pk, vk)$. Given public parameters $pp$ as input, output proving and verification keys $pk$ and $vk$. 
    \par\noindent $\textit{Prove}(pk,x,w) \rightarrow \pi$. Given the proving key $pk$, the public input $x$, and the witness $w$, output a proof $\pi$.
    \par\noindent $\textit{Verify}(vk,x, \pi)\rightarrow \{0,1\}$. Given the verification key $vk$, the public input $x$, and the proof $\pi$, output $1$ if the proof is valid and $0$ otherwise.

A \ZK-SNARK satisfies the following security properties:

\noindent \textbf{Knowledge Soundness. } A dishonest prover cannot convince the verifier of an invalid statement, except with negligible probability.

\par\noindent \textbf{Perfect Completeness. }
An honest prover can always convince the verifier of the veracity of a valid statement.

\par\noindent \textbf{Zero Knowledge. }
The proof $\pi$ reveals nothing about the witness $w$, beyond its existence.


For an in-depth introduction of \ZKPs and \SNARKs from a theoretical perspective, we refer the reader to~\cite{thaler2022proofs}.

\section{System and Threat Models for SNARKs}
\label{sec:model-threats}

We introduce a four-layer system model, showing how \SNARKs are implemented in practice.
Based on our system model, we provide a holistic threat model defining \SNARK vulnerabilities, adversaries, and their potential impact.

\begin{figure}[!t]
    \centering
    \includegraphics[width=0.5\textwidth]{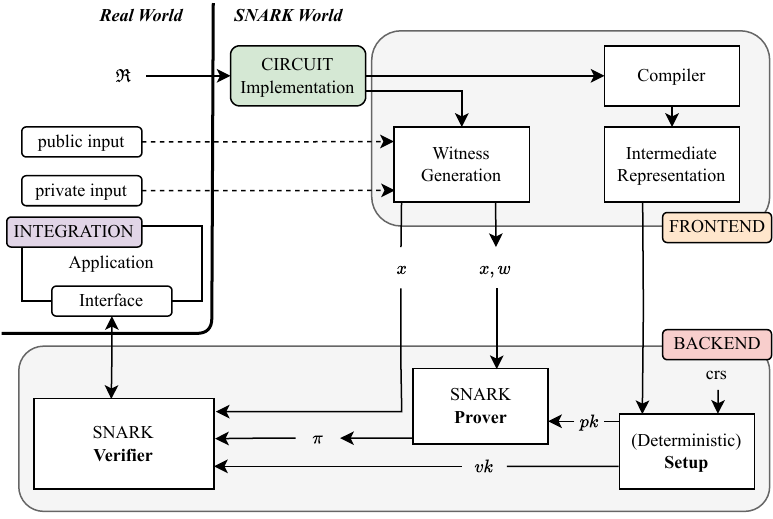}
    \caption{System model of an application based on \SNARKs.
        \extended{Note that witness generation can sometimes be produced by the compiler, while at other times it may be integrated into the backend.}
    }
    \label{figure:snark}
\end{figure}

\subsection{System Model}
\label{sec:model}

Figure~\ref{figure:snark} depicts the system architecture of an application based on \SNARKs for circuit satisfiability, i.e., argument systems that let the prover show that given a public input $x$, it knows a witness $w$ such that the circuit satisfies $\mathcal{C}(x,w) = y$. 
In our system model, we refer to four distinct layers. 
First, the developer specifies the circuit in the \textit{\CircuitLayer}, according to the specification of the statement to be proven.
The \textit{\FrontendLayer} enables the compilation of the circuit to a \SNARK-friendly representation.
The \textit{\BackendLayer} consumes this representation, and provides the concrete implementation of the proof system.
The \textit{\IntegrationLayer} is the application logic that interacts with the proof system.
We proceed to introduce the key components in each layer.

\par\medskip \noindent \textsc{\emph{(i)} \CircuitLayer. } 
\SNARKs targeting circuit satisfiability require the computational statement to be represented as an arithmetic circuit.
Most \SNARKs work over elements in a finite field $\mathbb{F}_p$, and hence each wire in the arithmetic circuit is represented as an element in $\mathbb{F}_p$. 
Note, that there are also \SNARKs employing Ring arithmetic~\cite{ganesh2023rinocchio}, with the purpose of achieving more efficient operations over $\mathbb{Z}_{2^{64}}$ to match native execution on standard CPUs. 
When developing a \ZK system, there are several considerations at different levels of the stack that must be taken into account by developers: \emph{(i)} determining what inputs and outputs of the computation are exposed publicly, \emph{(ii)} encoding the application logic as an arithmetic circuit in $\mathbb{F}_p$, and
\emph{(iii)} specifying how applications should compose with the circuit (e.g. by ensuring that public inputs are well-formed before verifying a proof).
Specifying circuits correctly is unintuitive, as arithmetic circuits do not natively support non-arithmetic operations.
Moreover, developers face challenges in translating conventional programming logic into circuit formats while simultaneously addressing the dual requirements of value assignment for witnesses and applying constraints to check the validity of solutions. The prover is responsible for assigning these values (i.e., witness assignments in the circuit), but the verifier has to check that these assignments adhere to the constraints (i.e., constraints in the circuit). Crucially, developers must remember that a malicious prover might manipulate witness values, bypassing circuit logic. Therefore, it is vital to rigorously constrain the witness within the circuit to only validate legitimate solutions.
If the constraints are not sound, then there is the risk of exploitation (see Section~\ref{sec:circuits}).
For example, expressing $X \neq 0$ is trivial in a ``normal'' programming language, whereas encoding it in a custom circuit is non-trivial.
By Fermat's Little Theorem, one can check $X^{p-1}=1$, but this would cost $O(log p)$ constraints~\cite{ozdemir2022circ}. 
An alternative is to have the prover provide a \textit{hint} $H = X^{-1}$, the multiplicative inverse of $X$, to check $X \cdot H = 1$.
Since all nonzero elements in $\mathbb{F}_p$ have a multiplicative inverse, this constrains $X$ to be nonzero by deduction~\cite{kosba2018xjsnark}.
We provide a concrete example of a circuit leveraging a hint in gnark for efficiently computing the square root in Figure~\ref{fig:golang_code}.
In practice, developer tools employ differing approaches to ease circuit specification in high-level programming interfaces:

\input{figures/codeListing}

    \par\medskip\noindent \textbf{Circuit Domain Specific Languages. }
    Domain Specific Languages (DSLs) are specialized programming languages designed to address specific problem domains. In the case of \SNARKs, they offer a tailored syntax to efficiently express constraints in an arithmetic circuit~\cite{parno2016pinocchio,eberhardt2018zokrates,chin2021leo,ozdemir2022circ,belles2022circom,amin2023lurk}.
    Learning a \DSL might be challenging, especially for developers not familiar with \SNARKs.
    \par\medskip\noindent \textbf{Circuit Embedded Domain Specific Languages. } 
    An Embedded Domain Specific Language (eDSL) is a type of domain-specific language that is embedded within a host general-purpose programming language. For developing \SNARKs, several eDSLs have emerged in recent years, embedded as libraries in Golang~\cite{gnark}, Rust~\cite{halo2, halo2_ce, plonky2}, and TypeScript~\cite{mina2021o1js}.
    eDSLs can seamlessly interact with other code written in the host language, allowing easy integration with existing libraries. At the same time, they require developers to actively distinguish between in-circuit and out-of-circuit operations, which requires domain expertise to ensure correct implementation.
    \par\medskip\noindent \textbf{ZK Virtual Machines. }
    Circuit DSLs and eDSLs allow developers to specify circuits in a manner similar to \textit{application-specific integrated circuits} (ASIC). Zero-Knowledge Virtual Machines (\ZK-VMs) follow a different programming model, where the arithmetic circuit represents the loop of fetching instructions from memory and successively executing them (similar to the fetch-decode-execute cycle as observed in a general-purpose CPU). Common \ZK-VMs target existing Instruction Set Architectures~\cite{scroll2023, polygon2023, era2023,bruestle2023riscZeroZkVM, arun2023jolt}.
    For example, \ZK-EVMs target the primitive instructions (i.e., ``opcodes'') of the Ethereum Virtual Machine~\cite{scroll2023, polygon2023, era2023}, and RISC Zero~\cite{bruestle2023riscZeroZkVM} and Jolt~\cite{arun2023jolt} operate on the RISC-V instruction set.
    Similarly, there are \ZK-VMs for custom ISAs that are optimized for proving in a \SNARK~\cite{polygon2023miden, goldberg2021cairo}. Some \ZK-VMs, such as the Cairo \ZK-VM~\cite{goldberg2021cairo}, provide a DSL to specify programs for the \ZK-VM.
    \ZK-VMs can be beneficial as they operate on existing instruction sets and can leverage existing tooling. 
    Writing circuits directly, while more error-prone, can be more efficient due to access to low-level optimizations. 

\par\medskip \noindent \textsc{\emph{(ii)} \FrontendLayer: } 
The frontend of a \SNARK for circuit-satisfiability compiles the high-level program written by the developer to a representation (i.e., \textit{arithmetization}) which is amenable for the proof system in a backend. Frontends are agnostic to proof systems, as in they can provide a compilation of high-level programs to differing arithmetizations. At runtime, the frontend assigns all intermediate wires in the circuit in order to generate the public and private parts of the \textit{witness}.
We introduce the components in detail as follows:

    \par\medskip\noindent \textbf{Arithmetization. }
    An arithmetization is a representation of the constraint system in a mathematical form, typically a set of algebraic equations, that can be efficiently processed by the argument system, i.e., the backend of a \SNARK.
    Common arithmetizations include rank-1 constraint systems (R1CS)~\cite{gennaro2013quadratic}, variations of Plonkish arithmetization~\cite{gabizon2019plonk}, Arithmetic Intermediate Representations (AIR)~\cite{ben2019scalable} and Customizable Constraint Systems (CCS)~\cite{setty2023customizable}.
    Different arithmetizations result in a different cost profile for the respective proof system that relies on them, and notably, they introduce different limitations with respect to how the circuit is defined. 
    For example, R1CS does not support constraints that have a polynomial degree larger than two, and AIR requires circuits to be uniform.
    Notably, Setty et al. provided CCS, a constraint system that generalizes Plonkish, AIR, and R1CS arithmetizations~\cite{setty2023customizable}.
    \par\medskip\noindent \textbf{Circuit Compiler. }
    The circuit compiler in a \SNARK implementation compiles the circuit specified in a high-level \DSL or eDSL to the respective arithmetization. 
    Note, that not all compilers target the compilation of high-level languages to specific arithmetizations.
    For example, high-level programs can first be compiled to VampIR~\footnote{\url{https://github.com/anoma/vamp-ir}} and ACIR~\footnote{\url{https://github.com/noir-lang/acvm}}, which are intermediate representations that allow compilation to, e.g., R1CS or some form of Plonkish arithmetization and aim to ease the support of multiple backends. 
    Similarly, zkLLVM provides a compiler from the LLVM IR to different arithmetizations, allowing users to prove the execution of, e.g., native Rust, Golang, or C++~\footnote{\url{https://github.com/NilFoundation/zkLLVM}}.
    \par\medskip\noindent \textbf{Witness Generation. }
    The main task of the witness generator is to calculate the intermediate wires for a given circuit $\mathcal{C}$, given the assignment of public and private inputs.
    Note that the concept of a separate witness generator is not necessarily implemented by every frontend. 
    Often, the circuit code can be executed to either produce constraints at compile time (e.g., the line highlighted in grey in Figure~\ref{fig:golang_code}) or to produce witness values at runtime/proving (e.g., the hint in Figure~\ref{fig:golang_code}).
    When an auxiliary ``witness generator'' exists, instructions are generated on how to fill a symbolic witness table and the witness generator is a binary generated at compile time.

\par\medskip\noindent \textsc{\emph{(iii)} \BackendLayer. } 
Given the arithmetic circuit and witness, the backend specifies the algorithmic implementation of proof systems (e.g., Groth16~\cite{groth2016size}, Plonk~\cite{gabizon2019plonk}, Stark~\cite{ben2019scalable},  Marlin~\cite{chiesa2020marlin}).
At a high level, the backend algorithms follow the API as in Section~\ref{section:background}. 


    \par\medskip\noindent \textbf{Setup. }
    In pre-processing \SNARKs, the \textit{setup} algorithm generates the prover and verifier keys ($pk$ and $vk$) by encoding the circuit relation, with the primary goal of succinct verification and optimized proving.
    However, different \SNARKs have vastly differing properties with regard to trust assumptions, which in turn also impact their performance.
    Some \SNARKs require a \textit{trusted setup}, i.e., a randomized setup phase that introduces a trapdoor 
    either per circuit (i.e., non-universal)~\cite{groth2016size} or in a universal setup phase~\cite{gabizon2019plonk}. 
    The \textit{randomized} part of the setup involves public parameters and random input to generate a common reference string (CRS)~\cite{nikolaenko2022powers}, which can be used to derive $pk$ and $vk$ in the deterministic part of the setup for a concrete circuit instantiation.
    The trapdoor needs to be discarded after the setup proccess is done, as anyone who has knowledge of the trapdoor can forge proofs. \SNARKs that do not employ a trapdoor in the setup phase are commonly coined \textit{transparent}~\cite{ben2019scalable}. Note, that this introduces a trade-off in verifier efficiency~---~transparent \SNARKs only have poly-logarithmic-time verification as compared to constant-time for \SNARKs with trusted setup~\cite{thaler2022proofs}.
    
    \par\medskip\noindent \textbf{Prover \& Verifier. }
    The prover consumes the values generated by the frontend, where the circuit is represented in a specific arithmetization.
    At a high level, the prover of a pre-processing SNARK (aside from Groth16~\cite{groth2016size}) utilizes a \textit{polynomial commitment} to commit to the satisfying witness, and successively engages with the verifier in an interactive protocol to evaluate the correctness of assignments with regard to the circuit relation.
    This ``recipe'' leads to a \SNARK if the verifier is public-coin~\cite{pass2011composition}, and the prover can hash the transcript of the interactive proof to render the protocol \textit{non-interactive} (a process commonly denoted as applying the ``Fiat-Shamir heuristic'').
    Different proof systems choose different combinations of polynomial commitments and interactive proofs, resulting in differing properties for the respective target applications. 
    The most popular approaches include Plonk~\cite{gabizon2019plonk}, which employs a constant-round polynomial interactive oracle proof (IOP) with the KZG commitment~\cite{kate2010constant}. Halo2~\cite{halo2} uses the Plonk IOP with the Bulletproofs polynomial commitment\cite{bunz2018bulletproofs}.
    Another class of protocol combines a polynomial commitment based on Merkle hashes with a \textit{low-degree test}, resulting in a protocol that does not demand for a trusted setup or operations in elliptic curve groups~\cite{ben2018fast}, at the cost of non-constant verification time and a larger proof size~\cite{ernstberger2023zk}.
    Plonky2~\cite{plonky2} uses the FRI-based polynomial commitment scheme and combines it with the Plonk IOP.
    Groth16~\cite{groth2016size}, the \SNARK with the shortest proof size, is based on probabilistically checkable proofs (PCP)~\cite{gennaro2013quadratic} and pairings over elliptic curves for proof verification. For an in-depth taxonomy, we refer readers to~\cite{thaler2022proofs}.
    Some applications leverage one of the above \SNARKs without the \textit{zero-knowledge} property, solely relying on non-interactivity and succinctness (e.g zk-EVMs~\cite{scroll2023, polygon2023}). In most \SNARKs, zero-knowledge can be obtained cheaply with minor modifications.

\par\medskip\noindent \textsc{\emph{(iv)} \IntegrationLayer. } 
In our system model, we collectively term any application-specific implementation that doesn't directly relate to the circuit layer, frontend layer, or backend layer, the \textit{integration layer}. This includes source code interacting with any of the other layers, application logic that may impact the overall security of employing \SNARKs, as well as composition and aggregation of proofs.

    \par\medskip\noindent \textbf{Code Interacting with \SNARK Components.} 
    A \SNARK application requires code that interacts with the functions exposed by the prior layers. Typically, developers use Solidity, JavaScript, or any other language to perform API calls to the frontend or backend. For performance reasons, many applications reduce circuits to their minimal size to fulfill a specific functionality, outsourcing operations, like range-checks for certain input values, to native application code.
    \par\medskip\noindent \textbf{Complementary ZKP Logic.} 
    An application might employ complementary logic beyond the \SNARK itself to ensure its security properties. For example, nullifiers are commonly used as private values that ``nullify'' a specific object upon its use to ensure that specific operations are only executed once.
    Consider the use of nullifiers in Tornado Cash~\cite{pertsev2019tornado}. A user deposits a token in a smart contract and associates a secret nullifier with it. 
    When a user wants to withdraw, the contract checks that the nullifier has not been used before to prevent double-spending. The user additionally generates a \SNARK proof, which proves that the nullifier is associated with a deposit, without revealing the exact deposit to achieve anonymity.
    \par\medskip\noindent \textbf{Proof Delegation, Aggregation, Recursion \& Composition. } 
    In some applications, proofs can be too computationally expensive to be generated on end-user devices. To achieve proof generation on computationally restricted devices, one may delegate the proof generation to an untrusted party~\cite{liu2023pianist}. 
    Similarly, one may require a fast prover (linear in the size of the statement) \textit{and} a fast verifier (constant in the size of the circuit). In this case, common projects employ a recursive composition of proofs, i.e., they first leverage a \SNARK with a fast prover and successively prove the verification in a \SNARK with a fast verifier. 
    For example, Polygon \ZK-EVM~\cite{polygon2023} composes a FRI-based \SNARK with a Groth16 to obtain the succinct proof size and verification for cheap verification of proofs in a smart contract, whilst benefiting from the faster prover time and decreased size of the reference string required for Groth16. Note that in this case, the trust assumption reduces to the weakest component, i.e., Groth16 still requires a trusted setup.
    There are several other works that employ a similar strategy to obtain efficient \SNARK constructions through aggregation or recursive composition~\cite{kothapalli2022nova, gailly2022snarkpack, xie2022zkbridge, rathee2022zebra, xie2022orion, el2022families}.

\subsection{Threat Model Taxonomy}


We consider any usage of a \SNARK that violates the intended behavior of one of the previously discussed layers as a \SNARK vulnerability. In this section, we first define the scope of vulnerabilities at each layer, outline the roles an adversary can take that may lead to a specific adversarial impact throughout our system model.


\par\medskip \noindent \textbf{Scope of Vulnerabilities. } In the \emph{(i)} \textit{circuit layer}, vulnerabilities may result from coding mistakes on the implementation of circuits that can lead to having inconsistent or weak constraints that break the soundness and/or completeness of the system. 
A primary reason for this is mistranslation due to developers not being used to writing in a differing programming model that introduces surprising pitfalls.
In the circuit layer, we exclude logic bugs that are not necessarily associated with writing \SNARK circuits, and focus on issues that arise from the use of \SNARK software specifically.
In the \emph{(ii)}~\textit{frontend layer}, vulnerabilities may primarily arise due to bugs in compiling from high-level source code to a specific arithmetization.
In the \emph{(iii)} \textit{backend layer}, vulnerabilities may occur in the prover, where adversarial provers attempt proof forgery.
For \SNARKs requiring a trusted setup, applications might use an MPC-ceremony for generating the reference string trustlessly. Vulnerabilities in the MPC-ceremony are considered out of scope, as they do not directly pertain to \SNARKs.
Vulnerabilities in the \emph{(iv)} ~\textit{integration layer} resemble any issue in the software components that interact with a \SNARK.
For example, circuits can have implicit constraints that ought to be checked in the integration layer.
We exclude traditional vulnerabilities, such as reentrancy for smart contracts, as they are not unique to systems using \SNARKs.

\input{tables/adversarialKnowledge}

\par\medskip \noindent \textsc{\emph{(i)} Adversarial Roles:} 
Throughout this work, we assume that adversaries are rational agents aiming to attack systems utilizing \SNARKs.
We outline the knowledge an adversary can obtain given its role in our system model in Table~\ref{tab:adversarial_knowledge}:

\par\medskip\noindent \textit{R1 - Network Adversary. } A network adversary can observe the public values transmitted between participants in the \SNARK application. Hence, it can obtain knowledge about the public witness $x$, the public common reference string, and the proof $\pi$ sent from the prover to the verifier. A network adversary may attempt to exploit weak simulation extractability of certain proof systems that can render a proof malleable~\cite{groth16malleability}.

\par\medskip\noindent \textit{R2 - Adversarial User. } In some \SNARK applications, users are not involved in proof generation or proof verification at all. For example, \ZK-EVMs provide a service that leverages \SNARKs primarily to minimize transaction costs. In this case, an adversarial user has oracle access to the prover, i.e., it can submit an arbitrary number of public inputs (or private inputs, in case of privacy-preserving delegated proof generation~\cite{chiesa2023eos}), with the aim of successfully performing Denial-of-Service (DoS) attacks. Further, an adversarial user may exploit circuit issues, where the circuit is defined by the prover (i.e., the service provider) to whom the user delegates the proof. 


\par\medskip\noindent \textit{R3 - Adversarial Prover. } 
An adversarial prover has knowledge of all input values, including the verifier key, and may attempt to break the underlying cryptographic primitives or exploit misconfigurations in the setup phase. The adversarial prover can easily exploit soundness vulnerabilities in the verifier, such as under-constrained bugs discussed in Section~\ref{sec:circuits}. To mitigate this risk, systems, including ZK-rollups, may temporarily adopt permissioned provers, where only authorized entities can generate and submit proofs to the verifier (typically a smart contract acting as the verifier). This approach reduces the risk of adversaries exploiting soundness vulnerabilities, although it does not eliminate the possibility of the permissioned prover exploiting such vulnerabilities.

\par\medskip \noindent \textsc{\emph{(ii)} Vulnerability Impact:}  We categorize the impact of a vulnerability into the following categories:

    \par\smallskip\noindent \textit{I1 - Breaking Soundness. } The vulnerability allows a prover to convince a verifier of a false statement; that is, it allows the creation of a proof for an incorrect statement that is nonetheless accepted as valid.
    
    \par\smallskip\noindent \textit{I2 - Breaking Completeness. } The vulnerability allows a prover to submit proofs of a true statement that leads to an invalid verification by a verifier. Further, it can be the case that valid proofs are rejected by an honest verifier.
    \par\smallskip\noindent \textit{I3 - Breaking Zero-Knowledge.} 
    The vulnerability allows an adversary to break the zero-knowledge property, i.e., it allows information leakage about the private witnesses.
    If an adversary exploits such a vulnerability, they could gain access to sensitive data, such as secret keys or private inputs, leading to privacy violations and potentially further exploits based on the information gained.


\input{tables/sources_table}

\input{tables/impact_table}

\section{Methodology}

Exploring the security vulnerabilities across the entire \SNARK stack presents a complex challenge for several reasons. Firstly, the unique programming model required for developing \SNARK circuits means that a significant number of potential security issues are difficult to identify and understand. 
Secondly, the tools used for \SNARK development are themselves non-standardized and heterogeneous, each offering different interfaces. 
The relative novelty of \SNARK technology contributes to a lack of comprehensive documentation and standards, further complicating the analysis of these tools. Moreover, instances of vulnerabilities being actively exploited are rare. There are no incidents of blackhat attacks publicly disclosed related to \SNARKs. Furthermore, in applications that leverage the zero-knowledge property of \SNARKs, it can be especially challenging to determine if an attack has occurred due to the privacy-preserving nature of these systems.
To cope with these challenges, we apply the following methodology.

\point{Analyzed \SNARK Implementations} 
Our examination focuses on widely deployed \SNARK systems, including \ZK-rollups for blockchain scalability (e.g., zkSync Era~\cite{era2023}, Polygon \ZK-EVM~\cite{polygon2023} Scroll~\cite{scroll2023}),\footnote{Notably, \ZK-rollups have more than 1B USD in accumulated value locked in them according to L2BEAT~\url{https://l2beat.com/scaling/summary}} privacy-centric blockchains (e.g., Zcash~\cite{sasson2014zerocash}, Aztec~\cite{aztec2023}, and decentralized privacy applications (e.g., Tornado Cash~\cite{pertsev2019tornado}). We also assess key circuit libraries and \SNARK frameworks (e.g., Circom and halo2)~\cite{ernstberger2023zk}. Our analysis further extends to inspecting the underlying proof systems for known bugs. This comprehensive approach ensures that we cover a broad spectrum of real-world \SNARK applications and the various layers of technology they rely on.

\point{Data Sources}
In our comprehensive study of security vulnerabilities related to \SNARK systems, we have utilized an extensive range of data sources to ensure a thorough analysis. Our primary source includes critical and high-impact vulnerabilities identified in~\totalaudits~security audit reports of systems employing \SNARKs. Among these,~\totalzkaudits~reports specifically highlighted vulnerabilities directly related to \SNARKs, while the remaining reports detailed other types of vulnerabilities, such as common smart contract deficits. Additionally, we have incorporated vulnerabilities from public bug bounty programs and vulnerability disclosures.
Our search for disclosures included those from prominent projects like Zcash, Aztec Connect, and TornadoCash, disclosures by security firms and researchers specializing in \ZKPs, and comprehensive web searches targeting \ZKP vulnerabilities. We reviewed Web3 bug bounty programs like Immunefi, though we found no \ZKP-related disclosures there. Additionally, we consulted closely with top audit firms focusing on \ZKPs to ensure we didn't miss any critical disclosures.
These sources provide real-world examples of how vulnerabilities have impacted operational systems as well as the measures taken to address them. Furthermore, we delved into bug trackers of popular GitHub projects related to \SNARKs. Our focus here has been specifically on security-related bugs, filtering out non-security-related issues. 
Table~\ref{tab:sources} in our paper depicts these varied data sources, collectively forming the foundation of our study of \SNARK security.


\point{Classification}
We classify vulnerabilities by determining the layer that they occur in with respect to our system model (cf. Section~\ref{sec:model}). 
Successively, we asses the vulnerability type (e.g., over-constrained), identify its root cause, and assess whether it impacts soundness, completeness, or zero-knowledge when exploited (cf. Table~\ref{tab:impact}).
We provide a full classification of all vulnerabilities as an open-source dataset as a reference for practitioners.\footnote{\url{https://docs.google.com/spreadsheets/d/1E97ulMufitGSKo_Dy09KYGv-aBcLPXtlN5QUpwyv66A}}

\extended{
 \point{Scope}
 The systems we analyzed might also be susceptible to vulnerabilities unrelated to \SNARKs, such as reentrancy issues in Solidity smart contracts within the integration layer~\cite{atzei2017survey}, type checker bugs in the frontend layer's compilers~\cite{chaliasos2021well}, or traditional supply chain attacks~\cite{ladisa2023sok} that can affect the circuit layer. 
 However, We deliberately targeted \SNARK-specific vulnerabilities to guide future research on enhancing \SNARKs security.
 We deliberately focused on \SNARK-specific vulnerabilities to identify future research directions for improving \SNARKs security. 
Yet, addressing general security challenges and following established best practices remain essential for overall system integrity.
}

\point{Threats to validity}
Our investigation into vulnerabilities within \SNARK systems faces several challenges that could affect the validity of our findings. Firstly, the absence of proof-of-concept exploits for many bugs raises questions about their actual exploitability. We try to mitigate this by excluding non-security vulnerabilities from our dataset. 
Secondly, while our classification of vulnerabilities was rigorously reviewed by multiple researchers and iteratively refined, the possibility of misclassification cannot be entirely ruled out. 
Our reliance on publicly available reports means our analysis may be skewed towards more commonly detected vulnerabilities, potentially overlooking others that are less apparent but equally prevalent. 

\section{Circuit Issues}
\label{sec:circuits}

\input{tables/circuits.tex}

Vulnerabilities at the circuit layer represent the most prevalent threat to systems using \SNARKs (c.f. \ Table~\ref{tab:sources}). 
The primary challenge for developers lies in adapting to a different level of abstraction, coupled with the need to optimize circuits for efficiency, as they significantly influence the cost of using a \SNARK.
This section initially highlights the three primary vulnerabilities encountered in the circuit layer, followed by outlining the root causes that lead to these vulnerabilities (see Table~\ref{tab:circuits}). In total,~\circuitsoundness~circuit vulnerabilities led to soundness issues and~\circuitcompleteness~led to completeness issues (c.f.\ Table~\ref{tab:impact}).
\extended{We present detailed examples of vulnerabilities and their root causes in Appendix~\ref{app:circuits}.}

\subsection{Vulnerabilities}

\vuln{Under-Constrained}
The most frequent vulnerability in \ZK circuits arises from insufficient constraints. This deficiency causes the verifier to mistakenly accept invalid proofs, thus undermining the system's soundness or completeness. 

\vuln{Over-Constrained}
Although less common than under-constrained issues, circuits can be over-constrained, leading to the rejection of valid witnesses by honest provers or benign proofs from honest verifiers. This issue stems from extra constraints in the circuit, where legitimate solutions cannot be proven or verified, leading to DoS issues.
Nevertheless, over-constrained bugs should not be confused with redundant constraints that add no additional value but do not lead to any issues other than introducing computational inefficiencies.

\vuln{Computational/Hints Error}
Occurs when the computational part of a circuit is erroneous, often leading to completeness issues where for correct inputs, the witness generation either fails or produces wrong results.
Note that completeness issues can often be transient, meaning that you can fix the underlying issue without having to update the circuit and recompute the prover and/or the verifier keys. Computational issues may also result in soundness issues if the constraints are applied using the same erroneous logic.

\subsection{Root Causes}

\rootcause{Assigned but not Constrained}
A frequent issue in \ZK circuit design lies in distinguishing between assignments and constraints. While constraints are mathematical equations that must be satisfied for any given witness for the proof to be valid, assignments allocate values to variables during the witness generation process. 
The problem arises when variables are assigned values based on correct computations but lack corresponding constraints. This oversight can lead the verifier to erroneously accept any value for these variables. 

\rootcause{Missing Input Constraints}
Developers sometimes neglect to apply constraints on input variables in reusable circuits. This omission occurs either (i) unintentionally or (ii) because they anticipate these constraints will be enforced at a different interface level (i.e., caller circuit or integration layer), thus omitting the constraints for optimization reasons. However, the absence of clear documentation or fully constrained inputs in these circuits can lead to severe vulnerabilities. 
Note that this issue is particularly common in low-level circuit \DSLs, such as Circom, which lack user-defined types.

\rootcause{Unsafe Reuse of Circuit}
In \ZK circuit design, particularly when using \DSLs like Circom, the practice of reusing circuits (such as templates in Circom or gadgets in halo2) can introduce vulnerabilities if not handled correctly. This primarily occurs in two scenarios:
\emph{(i)} \textit{Implicit Constraints in Sub-Circuits}, occurred when circuits are reused without appropriately constraining their inputs or outputs based on the assumption that the user will apply these constraints on call-site.
\emph{(ii)} \textit{Insecure Circuits Instantiation} when circuits are meant to be used for specific setups (e.g., specific curves). 
An example is the Sign template from circomlib, which was designed solely for the BN254 curve field.\footnote{\url{https://github.com/iden3/circomlib/blob/master/circuits/sign.circom\#L23}}

\rootcause{Wrong translation of logic into constraints}
Translating computations for \ZK circuits presents a unique set of challenges, primarily due to the distinct programming model of \ZK circuits compared to traditional CPU-targeted code. A significant issue arises when translating logic that involves types and operations available in conventional programming languages but are either absent or must be re-implemented in \ZK frameworks (e.g., fixed-point arithmetic). 
This often leads to the need for creative but error-prone solutions, such as using multiplexers for conditional logic. 
Developers might inadvertently omit essential constraints or simplify the logic to reduce the number of constraints, potentially missing critical corner cases. This can leave variables under-constrained, allowing them to accept multiple or any values under certain conditions, thus deviating from the developer's intent and introducing vulnerabilities. 

\rootcause{Incorrect Custom Gates}
In implementations following the TurboPLONK model, such as Halo2, circuit constraints are defined by using custom gates applied to specific rows and cells of a table that constitute the Plonkish representation of the circuit.
However, this approach can lead to bugs when custom gates are incorrectly handled. Errors may arise from inaccurately determining the appropriate offsets, resulting in misalignment with the intended behavior of the circuit. 

\rootcause{Out-of-Circuit Computation Not Being Constrained}
Out-of-circuit computation refers to computations within the code that do not directly impact the witness generation yet play a crucial role in the overall functionality. In \DSLs like Circom, certain functions operate outside the circuit logic.
Similarly, in \eDSLs, standard code (e.g., vanilla Rust) is used for various computations that do not affect witness generation.
For instance, computations like division, are typically performed out-of-circuit to optimize circuit efficiency. This method involves executing the computation externally and then witnessing the result back into the circuit where it is constrained to ensure correctness.
Issues arise when these out-of-circuit computations lead to missing assignments or when constraints necessary for the circuit's integrity are overlooked. A specific manifestation of this issue is the \textit{boomerang issue}, where a variable, initially constrained within the circuit, is temporarily moved out-of-circuit and then reintegrated without the necessary reapplication of constraints. 

\rootcause{Arithmetic Field Error}
Working with field arithmetic in \ZK circuits can be challenging, especially as developers might overlook the nuances of computations within a finite field. The most common issues in this context are arithmetic overflows and underflows. 
We categorize the primary types of overflows and underflows in \ZK circuits as follows: (i) \textit{Native Field Arithmetic Over/Underflow}, occurs when circuit computations exceed the finite field's limits, causing values to loop back within the field's range due to modulo arithmetic. (ii) \textit{Overflows in Transformed Formats}, risks of overflows arise when numbers are transformed into bit representations for specific operations (e.g., range checks) or for emulating non-native arithmetics like fixed-point arithmetics. This is particularly problematic when multiple bit representations remain within the field's overflow limits, leading to under-constrained vulnerabilities. 

\rootcause{Bad Circuit/Protocol Design}
Circuit design issues in \SNARKs often stem from fundamental flaws in how circuits are conceptualized, potentially leading to unintended behaviors or the violation of protocol properties. A common manifestation of this problem is the incorrect categorization of variables -- such as designating a variable as private when it should be public. These issues can significantly impact the functionality and security of the protocol. 


\rootcause{Other Programming Errors}
This category encompasses a range of common programming errors that do not fit neatly into other specific vulnerability categories but still have significant implications for the integrity of \SNARK systems. These errors include, but are not limited to, API misuse, incorrect indexing in arrays, and logical errors within the computational parts (i.e., witness generation) of the circuit that are not directly related to constraint application.

\point{Countermeasures} 
To address circuit layer vulnerabilities in \SNARK systems, especially the common issue of under-constrained vulnerabilities, several straightforward yet effective strategies are recommended. Firstly, adding missing constraints, particularly range checks, is crucial to ensure the integrity and robustness of the circuits. In-depth documentation of circuit design and \SNARK system specifications can significantly aid in preventing misunderstandings and oversights during development and auditing. Additionally, adopting \DSLs, whenever possible, that support modern programming features, such as abstractions and basic types, enhances the development process by providing clearer guidelines and reducing the likelihood of errors. Lastly, the use of specialized security tools designed for \ZK circuits, as detailed in Section~\ref{sec:defenses}, can detect some vulnerabilities during the developing phase, further securing \SNARK systems against vulnerabilities.

\section{Integration Layer Issues}

\input{tables/application.tex}

Besides circuit-related problems, numerous vulnerabilities in systems employing \SNARKs originate from the integration layer. These vulnerabilities often stem from improper interactions between the application/system and the prover or verifier, or from design flaws within this layer that compromise the inherent properties of \SNARKs. Table~\ref{tab:applications} categorizes all analyzed bugs into distinct categories and root causes.
\extended{For examples of integration-specific vulnerabilities, refer to Appendix~\ref{app:integration}.}

\subsection{Vulnerabilities}

\vuln{Passing Unchecked Data}
This vulnerability manifests when implicit constraints on the public inputs expected by the circuit are not enforced by the application's verifier. It can result in both soundness and completeness issues. While unchecked data can be input in the circuit directly or indirectly (by hashing it), the failure to enforce these implicit constraints at the integration layer can compromise the integrity of the entire system.

\vuln{Proof Delegation Error}
In scenarios where proof generation is delegated to an untrusted prover, there's a risk of malicious activities. For example, a bad design could lead to the leaking of personal or secret data. This issue underlines the need for secure and trusted channels in proof delegation and decentralized proving services~\cite{garg2023mathsf}.

\vuln{Proof Composition Error}
This issue arises when the logic is distributed across multiple proofs, but the intended behavior is not adequately enforced by the verifier who is expected to glue parts of certain proofs with parts of others. The lack of coherence and coordination between different proofs can lead to undefined behaviors in the application.

\vuln{ZKP Complementary Logic Error}
This category encompasses vulnerabilities in the integration layer arising from the flawed implementation of logic that operates in conjunction with \ZKPs. An illustrative example of such a flaw is the poor management of protocol-specific mechanisms like nullifiers. Consider a privacy-preserving application like Tornado Cash (TC), where users aim to dissociate deposits from withdrawals. TC employs \ZKPs alongside a nullifier scheme; each deposit involves generating a nullifier and a secret, which, when hashed, are appended in a Merkle tree in TC.\footnote{For more information on how TC works we refer the reader to the official documentation https://docs.tornadoeth.cash/generals/how-does-tornado-cash-work} Upon withdrawal, the user submits the nullifier and employs a \SNARK to prove knowledge of the corresponding secret without revealing it, thus unlinking the withdrawal from the deposit. However, a critical check within the integration layer is required to ensure the unique use of each nullifier. Without this check, the system is vulnerable to exploitation, allowing a user to repeatedly withdraw funds, thus draining TC. This class of bugs, crucially, is not due to a flaw in the circuit's design or implementation but stems from an oversight in auxiliary mechanisms that are integral to the secure and correct application of \ZKPs.

\subsection{Root Causes}

\rootcause{Missing Validation Input}
This root cause involves inadequate validation checks on data before its input into the circuit or before computations that affect data used later by the circuit. It arises when the integration layer does not properly validate input data or when the circuit itself lacks necessary checks. This oversight can lead to the acceptance of inappropriate data, potentially getting an application stuck, or worse, compromising the circuit's computations.

\rootcause{Integration Design Error}
Design flaws in the integration code that lead to undesired behavior or break protocol properties, such as not checking for duplicate nullifiers, are critical. Furthermore, designing privacy-enhanced applications poses a significant challenge. It is not straightforward to create \ZK applications with privacy as a core property, and developers may inadvertently leak information without realizing it. This issue isn't about the proof system or constraints; it is about the inherent difficulty in designing an application that doesn't subtly leak data. This underscores the complexity and importance of carefully considering privacy aspects in the design phase of \ZK applications. Fixes often require a high-level redesign rather than a simple programming correction.


\section{Frontend/Backend Issues}

\input{tables/framework.tex}

The compilation of \ZK circuits (frontend), along with proving a witness and verifying it (backend), rely on the correctness of the underlying software that implements the compilation process as well as the algorithms that underpin the proof system. Table~\ref{tab:framework} provides an overview of vulnerabilities that can occur within these components. 
Vulnerabilities in either the frontend or backend layers pose a significant risk,
even if the circuits have been formally verified or if the proof system is theoretically secure; any defects in these layers could render the entire system insecure for end-users. 

\subsection{Frontend Vulnerabilities}

\vuln{Incorrect Constraint Compilation}
This type of vulnerability emerges from deficiencies in the compilation phase, specifically in the enforcement of constraints defined in the \ZK circuit. Specifically, such issues may stem from the arithmetization process or overly aggressive optimization routines that unintentionally excise vital constraints. Any missteps in these processes can result in an inaccurate or incomplete translation of the original constraints at the circuit layer. Consequently, this could permit a verifier to erroneously accept an invalid proof or, conversely, to reject a legitimate one. At its core, this vulnerability mirrors a classical compiler bug, wherein the compiler erroneously interprets high-level \DSL code.

\vuln{Witness Generation Error}
This issue arises during the witness generation phase, where errors in the compilation can produce invalid witnesses or cause crashes, potentially leading to a denial of service. Typically, such errors occur due to misinterpretation of the circuit code, or due to the generation of bogus witness generators in target languages. 

\subsection{Backend Vulnerabilities}

\vuln{Setup Error}
This vulnerability arises during the setup phase, where the generation of public parameters occurs. Incorrect or easily compromised parameters can significantly undermine the system's integrity.

\vuln{Prover Error}
Issues within the prover of the proof system fall into this category. These vulnerabilities can lead to the prover mistakenly rejecting valid witnesses, accepting invalid ones, or breaking the zero-knowledge property. 

\vuln{Unsafe Verifier}
This category includes vulnerabilities resulting from inadequate checks on the verifier's inputs, missing checks during verification, or bogus mathematical operations,  risking the proof system's integrity. Vulnerabilities can emerge regardless of whether the verifier is manually implemented or generated by a framework.

\subsection{Root Causes}
Frontend and backend vulnerabilities can stem from a variety of root causes, ranging from basic programming errors to configuration issues or deviations from the specifications of proof systems. 
These can be as simple as missing validation checks for proofs or as complex as informational leaks that expose sensitive data. Additionally, the use of poor-quality or predictable randomness sources can compromise crucial \ZKP properties. 
Adherence to proof system specifications is critical in preventing vulnerabilities. A common issue is the incorrect implementation of the Fiat-Shamir transformation, where a critical component is omitted during the hashing process of the transcript. 


\input{tables/defenses.tex}
\section{Defenses For SNARKs}
\label{sec:defenses}

This section presents an overview of defense mechanisms aimed at mitigating the \SNARK vulnerabilities detailed in the previous sections. Table~\ref{tab:snark-defenses} compiles, to our best knowledge, all publications and tools associated with \SNARK security. A filled bullet in the table indicates that the referenced technique provides at least one defensive measure against the issues indicated by the corresponding row's header. Notably, in addition to \SNARK-specific strategies, conventional security tools (e.g., fuzzing) have been utilized in audits, although they typically fall short in preventing most vulnerabilities due to the oracle problem~\cite{barr2014oracle}. For example, fuzzing might be useful in detecting completeness bugs, but it often falls short of finding soundness issues.
In the following, we delve into the current defensive measures for each layer, suggest future steps for preventing \SNARK vulnerabilities, and review the tools employed in the audits we examined.

\point{Circuit Layer Defenses}
The circuit layer emerges as the most prominent layer for \SNARK vulnerabilities, leading to the development of various techniques like static analysis and symbolic execution, particularly targeting under-constrained bugs. Tools such as Circomspect~\cite{circomspect} and ZKAP~\cite{wen2023practical} employ static analysis using predefined rules to identify under-constrained issues in Circom circuits. Similarly, Korrect~\cite{soureshjani2023automated} utilizes static analysis and SMT solvers to spot common pitfalls in Halo2 circuits. SNARKProbe~\cite{fansnarkprobe} leverages fuzzing and SMT solvers to test circuits written using R1CS-based libraries. Picus~\cite{pailoor2023automated}, on the other hand, adopts a more advanced symbolic execution approach to verify that Circom circuits are not under-constrained. CIVER~\cite{isabel2024scalable} uses a modular technique based on the application of transformation and deduction rules to verify properties of Circom circuits using pre- and post-conditions. In a different approach, DSLs like Coda~\cite{liu2023pianist} and Leo~\cite{coglio2023formal,coglio2023compositional} support formal verification of circuits, aiding developers in creating more secure circuits.

Despite these initial strides in tool development for \ZK circuits, significant limitations remain. Tools relying on SMT solvers, like Picus and Korrect, face challenges due to limited support and efficiency in handling finite field arithmetic, leading to performance bottlenecks~\cite{hader2023smt,ozdemir2023satisfiability}. Static analysis tools are often restricted in the range of vulnerabilities they can detect and tend to be specific to certain languages. CirC~\cite{ozdemir2022circ} compiler represents a step towards language-agnostic tooling, allowing various DSLs to compile into an Intermediate Representation, which can then be analyzed for potential vulnerabilities. Another promising direction is the creation of more secure DSLs, like Aleo~\cite{chin2021leo}, noname~\footnote{\url{https://github.com/zksecurity/noname}}, and Noir~\footnote{\url{https://aztec.network/aztec-nr/}}, or eDSLs such as o1js~\cite{mina2021o1js}, which incorporate strong typing and improved abstractions to prevent common issues.

Additionally, differential testing~\cite{mckeeman1998differential} against a reference implementation could be a viable method to identify computational issues. We anticipate that proper compilation of DSL to an IR could enable more general static analysis to uncover \SNARK vulnerabilities in a DSL-agnostic way. Furthermore, applying hardening techniques to automatically patch circuits with missing range checks could enhance security. Complementary to existing approaches, compositional verification techniques could be applied to verify \ZK circuits. Lastly, advanced fuzzing techniques that generate slightly incorrect witnesses might prove effective in detecting under-constrained vulnerabilities, while providing counterexamples. Finally, there is a growing demand for more advanced testing frameworks in the field. Such frameworks would greatly assist developers in creating their own unit and integration tests, which could be instrumental in identifying and rectifying many easily detectable vulnerabilities prior to deployment.

\point{Integration Layer Defenses}
Table~\ref{tab:snark-defenses} highlights a notable gap: there are currently no specific defenses developed for vulnerabilities in the integration layer of \SNARK systems. However, traditional security methods like property-based fuzzing or formal verification could be applicable if there's a comprehensive specification detailing the expected system behavior and its interactions with \SNARK components. A promising direction for future development is the creation of security frameworks designed to rigorously test the integration between client-side code (such as Solidity and JavaScript) and circuit code. We suggest exploring multi-layer testing and verification techniques targeting both these components simultaneously. An example could be a combined testing of Circom and Solidity applications. Additionally, having a detailed specification of the entire system, including how different components interact, can significantly enhance the effectiveness of manual code inspection.

\point{Frontend/Backend Layers Defenses}
The infrastructure layer, comprising the frontend and backend of \SNARK systems, has largely been overlooked by security researchers. Ozdemir et al.~\cite{ozdemir2023bounded} acknowledged that a flaw in a \ZKP compiler could undermine the integrity of a \ZKP, and took initial steps to mitigate such risks by partially verifying a key compiler pass in the CirC compiler~\cite{ozdemir2022circ}. SNARKProve~\cite{fansnarkprobe} implements a fuzzing framework that can be configured with custom files (i.e., \textit{ideal files}) to test specific cryptographic properties of the backends and frontends. In its initial version, it has been successful in detecting issues in the Setup phase of the backend. Moreover, practitioners have employed fuzzing tools like AFL to identify bugs in these layers. However, these tools are somewhat limited in scope, typically only capable of detecting crashes, potentially missing more subtle but critical vulnerabilities such as miscompilations.

As the field progresses with more complex optimizations in the compilation stage being enabled and the introduction of more advanced proof systems, it is likely that more vulnerabilities will emerge at the infrastructure layer of \SNARKs. Consequently, there is a pressing need for the development of more sophisticated testing methodologies for the infrastructure layer. Overcoming the oracle problem and generating effective test cases are key challenges in this area. Insights from the extensive body of work on testing conventional compilers~\cite{yang2011finding,chen2013taming,le2014compiler,chaliasos2022finding} could be valuable in devising strategies for more effectively testing \SNARK compilers.

\point{Tools Applied on Audits}
Our analysis of the \empirical{$75$} audit reports we reviewed revealed that \SNARK-related tools were utilized in only $5$ instances. Specifically, Picus was employed in $4$ audits, while ZKAP, Ecne, and Circomspect were each used once. Interestingly, one audit incorporated differential testing against a reference implementation to identify computational issues in the circuits. Moreover, traditional security tools like AFL, Semgrep, and property testing were used in $10$ audits, primarily for detecting bugs at the circuit and infrastructure levels. This highlights a significant need for enhanced tooling for both the frontend and backend layers of \SNARK systems.

\point{Multi-Provers as a Defense Mechanism}
\SNARK proofs are often used in safety-critical contexts, yet, with the current state of technology, guaranteeing that a complex \SNARK system is bug-free and safe remains an elusive goal. The multi-prover design enhances safety by introducing redundancy at the proof system level. A multi-prover system utilizes multiple proof systems besides the primary \SNARK, such as alternative \SNARK proofs or trusted-execution environment (TEE) based proofs~\cite{cerdeira2020sok}, and primary proofs are accepted only if the secondary ones agree with it. This design trades off liveness with more safety, as a dispute will potentially bring the system to a halt. Such systems require a dispute resolution mechanism to handle potential disagreements. In practice, some \ZK-Rollup projects \cite{scroll2023}, arguably one of the most complex categories of \SNARK systems today, are already experimenting with multi-provers.\footnote{\url{https://scroll.io/blog/scaling-security}}

\section{Issues in Proof Systems}

While the focus of earlier sections was on implementation and design vulnerabilities across the four layers introduced in Section~\ref{sec:model}, vulnerabilities can also exist within the theoretical foundations of \SNARKs' proof systems. This includes the formal descriptions and security proofs of protocols. Such vulnerabilities are less common but carry significant implications, potentially affecting all implementations of a given proof system. For example, the ``Frozen Heart'' vulnerability within the Plonk proof system (c.f.\ Appendix~\ref{app:frozen}) led to standard implementations of Plonk being compromised. The discovery of such flaws could severely impact the \SNARK ecosystem, especially protocols that depend on these systems. Vulnerabilities in the proof system generally stem from two main sources: errors in the original proof system description, including missing or incorrect security proofs or incomplete descriptions that could lead developers to introduce significant vulnerabilities during implementation. 
\extended{We provide a detailed account of known vulnerabilities in Appendix~\ref{app:proof-system}.}

The primary method for evaluating proof systems currently is peer review, where researchers assess security proofs, supplemented by occasional manual audits. Beyond cryptanalysis, employing tools like EasyCrypt~\cite{barthe2012easycrypt} and other computer-aided cryptographic proof software~\cite{barbosa2021sok} offers a promising avenue for formalizing and verifying the security properties of zero-knowledge protocols, as exemplified by Firsov et al.'s work~\cite{firsov2023zero}. It is crucial for every proof system to be accompanied by exhaustive security proofs and to undergo rigorous review before production use to prevent potential bugs.

\point{Universal Composability}
The universal composability (UC) framework by Canetti~\cite{canetti2001universally} is widely considered as the ``gold standard'' for proving the security of cryptographic primitives. The reason for the popularity of the UC model is that it can guarantee strong security against adaptive adversaries and further allows for modular reusability of cryptographic primitives in greater, high-level protocol designs.
Arguing UC security for a \SNARK, or a \SNARK-based protocol, is non-trivial, as \SNARKs commonly use techniques that are not realizable in the UC model~\cite{groth2006simulation}, and result in \SNARKs that are not formally non-malleable.
Recent work aims to close this gap by studying compilers that render common \SNARKs UC-secure by encrypting the witness and including it in the argument, which results in an overhead that renders the \SNARK non-succinct due to the unbounded size of the witness~\cite{kosba2015c, abdolmaleki2023universally, abdolmaleki2020lift, baghery2021tiramisu}. Most recent work proves that initial \SNARK constructions~\cite{micali2000computationally, ben2016interactive} that \textit{do} utilize random oracles are UC-secure~\cite{chiesa2024zksnarks}.

In practice, we observe that UC security is often not considered, and non-UC-secure protocols, like Groth16~\cite{groth2016size}, are applied without further ado. However, we are not aware of any protocol that claims UC-security and successively got exploited due to erroneous analysis, i.e., an underspecified ideal functionality or incorrectly claiming that the real-life model can be simulated in the ideal process model.

\section{Discussion}
In the following, we extract insights on the current state of \SNARK security, highlight key findings, discuss their
implications and make recommendations for future research.

\noindent \textbf{1) Insight -- Under-constrained bugs pose a significant threat to \SNARK deployments.}  Under-constrained bugs emerge as the most prevalent vulnerability class within \ZK circuits. Unlike typical vulnerabilities, their root causes span from straightforward programming errors to challenges inherent in \SNARK \DSLs, including the complexity of translating logic to constraints efficiently. As developers navigate these peculiarities, ensuring circuits are thoroughly constrained remains crucial for maintaining system integrity and security.

\noindent \emph{RQ}: \textit{What tools and methodologies can be developed to better identify and mitigate under-constrained bugs in \ZK circuits? Which techniques are the most efficient in detecting such bugs?}

\noindent \textbf{2) Insight -- Soundness bugs affecting \SNARK verifiers lead on average to high severity bugs.} 
Particularly when soundness bugs are exploitable, typically due to under-constrained circuits, they can lead to significant security breaches. Their potential to compromise system security has led to suboptimal strategies, such as employing multi-provers and permissioned provers, alongside significant bug bounties, sometimes reaching up to \$500k for a single vulnerability.~\footnote{\url{https://aztec.network/blog/aztec-connect-postmortem/}} These measures underscore the critical nature of soundness in maintaining the trustworthiness of \SNARK-based systems.

\noindent \emph{RQ}: \textit{What are effective strategies to detect and prevent soundness bugs in \SNARKs? Can we formally verify the verifiers, as their scope is limited?}

\noindent \textbf{3) Insight -- Low-level circuit DSLs are easy-to-misuse leading to many vulnerabilities.}
Crafting efficient \ZK circuits often necessitates using low-level \ZK \DSLs such as Circom and Gnark, reminiscent of early assembly and C programming, where common high-level programming features such as abstractions and basic types are absent. This complexity not only steepens the learning curve for developers but also increases the likelihood of introducing vulnerabilities into the circuits. The prevalent use of such \DSLs highlights a regression to an era with bugs due to the absence of modern programming safeguards, underscoring an urgent need for more user-friendly DSLs that offer better abstractions and safety features, thereby mitigating the main shortcomings and reducing the vulnerability surface of \SNARK applications.

\noindent \emph{RQ}: \textit{How can more user-friendly \DSLs be designed to reduce the vulnerability surface of \SNARK applications?}

\noindent \textbf{4) Challenge -- The added complexity of \SNARKs present challenges for developers and auditors.}
The inherent complexity of \SNARKs introduces significant challenges for both developers and auditors, compounded by the abstraction levels of \ZK circuits and the low-level intricacies of most \DSLs. Developers often find themselves navigating the task of integrating critical cryptographic operations—such as digital signatures, commitment schemes, and Merkle trees—within the \ZK circuits. This combination of complex cryptographic code with the unique paradigm of \SNARKs places a considerable burden on ensuring accuracy and security, particularly when such code underpins the most vital or privacy-sensitive parts of an application. 
Hence, not only must developers acquire the technical depth of cryptographic programming within these new frameworks, but auditors must also adapt their methodologies to effectively scrutinize these sophisticated systems.

\noindent \emph{RQ}: \textit{What educational and tooling resources can assist developers and auditors in navigating the complexities of using  \SNARKs? How useful can specifications (formal or informal) be towards detecting vulnerabilities in systems using \ZKPs?}

\noindent \textbf{5) Insight -- Compiler and proof system implementation bugs can undermine major protocols.}
Infrastructure bugs in compilers and proof system implementations can critically undermine the security of major \SNARKs. Even when circuits are verified and audited, vulnerabilities in the underlying infrastructure or proof systems can jeopardize the entire application, emphasizing the importance of holistic testing approaches for the infrastructure layer of the \SNARKs. 

\noindent \emph{RQ}: \textit{How can the reliability and security of compilers and proof system implementations for \SNARKs be ensured? Which testing or verification techniques can be applied?}

\noindent \textbf{6) Challenge -- Preliminary security tools show promising results but also limitations.}
Recent developments in tools for securing \ZK circuits show promise but face scalability issues and are often limited to specific \DSLs or types of vulnerabilities. The complexity of \SNARK systems makes manual code inspection necessary, pointing to a significant need for better security tools and educational resources. This combination of advanced tooling and increased knowledge is crucial for improving the security of the \SNARK ecosystem.

\noindent \emph{RQ}: \textit{What improvements are needed in security tools to effectively scale and cover a broader range of vulnerabilities? What are the limitations of formal verification tools?}

\noindent \textbf{7) Insight -- Insecure Proof System Instantiation.}
The selection of cryptographic curves and fields for \SNARK instantiation is a critical decision that can significantly impact system security. Insecure choices can expose the system to potential brute force attacks similarly to other cryptographic protocols, underscoring the necessity of careful selection to ensure the overall security and integrity of \SNARK applications. 

\noindent \emph{RQ}: \textit{What is the security of \SNARKs instantiation given specific curves and fields? How many bits of security does each \SNARK have when using specific configurations?}

\noindent \textbf{8) Challenge -- \SNARK undetectable exploits.}
In privacy-preserving blockchains, such as ZCash, coins enter a privacy pool (a.k.a. shielded value pool) where they can be transferred without revealing the amount or recipient address. The transaction validation mechanism relies on \SNARKs and a soundness bug can enable the adversary to print infinite coins inside the privacy pool. Note that such attacks can remain undetected as long as, at any point, the amount of coins that exited the pool is smaller or equal to coins that have entered the pool previously. 
A partial defense against these attacks is \textit{turnstile enforcement}, adding a consensus check to reject blocks violating pool balance, ensuring exits don't exceed entries. This highlights the need for privacy-focused \SNARK systems with exploit detection and prevention mechanisms.

\noindent \emph{RQ}: \textit{How can we develop detection mechanisms or heuristics for exploits in privacy-preserving \SNARK systems?}

\section{Related Work}
\point{\SNARK Vulnerabilities}
This paper introduces a four-layer system model, defines threat models along with a detailed taxonomy of vulnerabilities and root causes across each layer in \SNARK systems. Prior works have identified specific vulnerabilities within \SNARKs; for example, Wen et al.~\cite{wen2023practical} highlighted common vulnerabilities in Circom circuits, while others~\cite{soureshjani2023automated,wang2022ecne,pailoor2023automated,fansnarkprobe,isabel2024scalable} have focused mainly on under-constrained vulnerabilities and proposed countermeasures. Ozdemir et al.~\cite{ozdemir2023bounded} shed light on potential issues during the \SNARK compilation phase. For security tools related to \SNARK vulnerabilities, Table~\ref{tab:snark-defenses} offers a comprehensive overview. Additionally, the community maintains a bug tracker dedicated to \ZKP-related vulnerabilities.\footnote{\url{https://github.com/0xPARC/zk-bug-tracker}} Our research complements these efforts by systematizing knowledge from the examination of~\totalvulns vulnerabilities and enriching the understanding of \SNARK security.

\point{Security of Integrity-Preserving Technologies}
Integrity-preserving computation encompasses a variety of technologies, each presenting unique security challenges. Similar to our work, Cerdeira et~al.~\cite{cerdeira2020sok} examined~$124$ CVEs within TEE-based systems and proposed a vulnerability taxonomy for TrustZone-assisted TEE Systems.
Blockchain smart contracts represent another avenue for integrity-preserving computation; Praitheeshan et~al.~\cite{praitheeshan2019security} identified common software and Ethereum smart contract vulnerabilities, focusing on issues prevalent at the smart contract layer.
Homoliak et~al.~\cite{homoliak2020security} introduced a multi-layered security model, systematically addressing vulnerabilities, threats, and countermeasures for blockchains. 
Atzei et~al.~\cite{atzei2017survey} explored Ethereum's security vulnerabilities, offering a classification of common programming pitfalls. Zhou et al.~\cite{zhou2023sok} developed a five-layer model and a comprehensive taxonomy of threat models to analyze and compare incidents in DeFi. Further, Chaliasos et al.~\cite{chaliasos2023smart} used the dataset from~\cite{zhou2023sok} to evaluate state-of-the-practice smart contract security tools against real-world vulnerabilities. In a similar way, future work could leverage our findings to assess the efficacy of emerging security tools for \SNARKs against the vulnerabilities detailed in our dataset.

\section{Conclusions}
In this work, we present comprehensive system and threat models for \SNARK systems' security, a detailed study, and a taxonomy of \totalvulns vulnerabilities, demonstrating that security breaches can affect every layer of systems employing \SNARKs, jeopardizing completeness, soundness, and zero-knowledge properties. Our work reveals the intricate and unique security challenges inherent to \SNARKs, indicating that defense mechanisms focusing on \SNARK security have significant limitations. By highlighting key insights and potential advancements in security practices, we underscore the urgency for continued research and enhanced defenses within the \SNARK ecosystem. As \SNARKs become increasingly pivotal in cryptographic applications, our study emphasizes the necessity for progressive and fortified security measures to ensure the robustness of these systems.

\section*{Acknowledgments}
We would like to thank our shepherd and the anonymous USENIX reviewers.
We also thank Shankara Pailoor, Alex Ozdemir, Jean-Philippe Aumasson, Oana Ciobotaru, and Justin Traglia for their insightful feedback.
Finally, we thank the Ethereum Foundation and 0xPARC for supporting this work.

\bibliographystyle{plain}
\bibliography{main}

\appendix

\section{Proof System Vulnerabilities}
\label{app:proof-system}

\subsection{Counterfeiting -- Setup Issue}
Ben-Sasson et al. \cite{ben2014succinct} presented a variant of Pinocchio \cite{parno2016pinocchio} with a shorter verification time and almost similar proving time. Given the similarity between the two works, the authors do not provide concrete proof of security, but rather, they reference Pinocchio's security proof \cite{parno2016pinocchio}. Despite the popularity and adoption of this work by major projects such as ZCash, the lack of security proof turned out to be a fatal choice, as critical soundness issues were discovered \cite{parno2015note, gabizon2019securityBCTV}.

The trusted setup procedure described in \cite{ben2014succinct} produces extra elements that are not used neither by the verifier or an honest prover, however they can be used to break the soundness of the proof system. These ``bypass" parameters could be recovered from the transcript of the trusted setup ceremony. In particular, the issue enables a malicious prover, to create a proof of knowledge for \textit{any} public input given a valid proof for \textit{some} public input. In the case of ZCash, this enabled an attacker to produce accepting yet invalid proofs of knowledge of membership witnesses for coins in the privacy pool, which is enough to create counterfeit new coins \cite{zcash2019counterfeiting}.

\subsection{Frozen Heart -- Insecure Fiat Shamir Transformation}
\label{app:frozen}
Fiat-Shamir (FS) transformation is a ubiquitous tool to compile public-coin interactive protocols into non-interactive ones, and many \SNARKs constructions rely on it. The idea is simple: in the interactive protocol, replace any instance of random challenge from the verifier (a.k.a. public-coin flip) with the cryptographically secure hash of \textit{the whole transcript} of the protocol up until that point; this way, the prover does not need to interact with the verifier anymore. The transformation hinges on the output of the hash function being pseudo-random. 

Note that if at any point a malicious prover does not like the challenge produced by Fiat-Shamir, his only option is to change his previous messages, but given that the hash function is cryptographically secure, he is unable to efficiently find a suitable change. However, invoking Fiat-Shamir with only a \textit{partial transcript} of the protocol can create serious soundness issues, referred to as \textit{Frozen Heart attacks} \cite{dao2023weakfiatshamir,bulletproofs2019frozen}. Intuitively, this enables a malicious prover to execute the steps of the interactive version of the proof system in an incorrect order, likely breaking the assumptions in the security proof. In particular, the issue enables a malicious prover, to create an \textit{invalid} proof of knowledge for \textit{some} public input given a valid proof for another public input.

\subsection{The Last Challenge Attack}
The Last Challenge Attack~\cite{ciobotaru2024the} specifically exploits a flaw in KZG-based SNARK implementations where the final FS transform challenge (i.e., the KZG batching challenge) is computed by a non-compliant verifier using only a part of the full communication transcript. In this scenario, a malicious prover can manipulate the proof process to produce a valid proof for a false statement. The attack occurs because the malicious prover can set to arbitrary values of its choice all public inputs as well as some of the components of the SNARK proof, and for the rest of the components, the prover can, with high probability, compute values that pass the non-compliant verification check. Consequently, the malicious prover can create an invalid proof for some public input by leveraging a valid proof for another public input. This can lead to disastrous consequences in practice.

While both the Last Challenge Attack and the Frozen Heart vulnerability exploit incorrect implementations of the FS transform, and both may allow for creating proofs of incorrect statements, they differ in specifics. The Frozen Heart vulnerability may occur if parts of the public input are omitted by the FS transform.
The Last Challenge Attack may occur if one omits parts of the transcript, other than the public input, in the computation of the final FS transform challenge. 

\subsection{Soundness and Malleability in Nova IVC}
Incrementally Verifiable Computation (IVC) is a cryptographic primitive that allows efficiently generating proofs for ongoing computation and produces succinctly verifiable proofs for any prefix of the computation. Nova \cite{kothapalli2022nova} is a proof system that utilizes folding to realize IVC. Folding is a procedure to aggregate two NP statements into a new one in a way that proves the new statement is enough to prove the two original ones. 

In Nova IVC, at each new round, the statement regarding the prefix of computation so far is folded with the statement regarding the final step; this folding has to be done in the circuit, which could involve non-native arithmetic operations that are relatively costly. One approach to deal with non-native arithmetic operations is to use cycles of elliptic curves, switching the circuit and curve at each step, so that almost all of the in-circuit arithmetic operations are native. The original implementation of Nova indeed used a 2-cycle of curves. This variation of the proof system was not fully specified nor proved to be secure in the original paper. Nguyen et al. \cite{Nguyen2023NovaCycle} showed that the implementation misses a critical check to verify the consistency of the inputs to the two circuits, breaking the soundness completely. Notably, \cite{Nguyen2023NovaCycle} proposes a fix for this issue and provides proof of security.

Moreover, \cite{Nguyen2023NovaCycle} shows that for non-deterministic computation steps, that is, computations that rely on auxiliary inputs (witness), Nova IVC proofs are malleable. That is, an attacker can create \textit{invalid} proofs of knowledge for \textit{some} public input given a valid proof for \textit{any} public input. The idea is that if the attacker can extract the witness for the prefix without the final step, then he can change the auxiliary input for the final step and derive a new proof that attests to the knowledge of auxiliary inputs in the whole chain of computation. 

\section{ZK Vulnerabilities Examples}
\label{app:examples}

\input{tables/examples_table.tex}

In Table~\ref{tab:examples}, we present four examples of vulnerabilities relevant to our study on the security of \SNARK systems. These examples demonstrate the range and complexity of possible security issues that could compromise the trustworthiness of systems based on \SNARKs. To enhance our analysis and offer a deeper understanding of these vulnerabilities, we further explore additional examples, including code snippets, focusing specifically on the circuit and integration layers. 

\subsection{Circuit Layer Vulnerability Examples}
\label{app:circuits}

\subsubsection{Underconstrained -- Assigned but Unconstrained}

\textbf{DSL:} Circom \\
\textbf{Project:} Circomlib \\
\textbf{Source:} \url{https://tornado-cash.medium.com/tornado-cash-got-hacked-by-us-b1e012a3c9a8} \\
\textbf{Fix:} \url{https://github.com/iden3/circomlib/pull/22/files}

\point{Circom background}
Circom enables low-level expression of computations akin to writing arithmetic circuits, while specifying quadratic, linear, or constant equations for constraining inputs and outputs. 
When compiling a Circom program, two essential artifacts are produced: a witness generation program and an encoding of the program in Rank-1 Constraint System (R1CS). 
Circom operates on ``signals,'' which are finite field elements (or arrays) variables defined with the \texttt{signal} keyword. 
These signals can be designated as inputs or outputs, while those without such annotations are treated as intermediate signals. 
The witness generator, created by the Circom compiler, calculates the values for all intermediate and output signals based on the input. 
Circom includes specific operators for computation (witness calculation) and constraint generation (R1CS creation). 
The operators \texttt{<--} and \texttt{-->} are employed for signal assignment, directing witness generation, whereas \texttt{===} is used for generating constraints; 
for instance, \texttt{x === y - 1} instructs the Circom compiler to produce a circuit accepting only when \texttt{x} equals \texttt{y - 1}. 
To streamline the process where witness generation and constraint logic overlap, Circom introduces \texttt{<==} and \texttt{==>} operators for simultaneous signal assignment and constraint, effectively merging these two aspects.

A common error in \DSLs like Circom, which provide distinct APIs for witness generation and constraints, is assigning a value to a witness without subsequently constraining it. 
This oversight introduced a vulnerability within Circomlib's MiMC hash template, leading to a critical vulnerability in Tornado Cash~\cite{gurkan2019tornadocash}, potentially allowing all tokens from its pools to be drained. 
Figure~\ref{fig:assigned_but_unconstrained} illustrates this flaw. The Circom template implements the \texttt{MiMC-2n/n} hash using a sponge construction.
In line $10$, \texttt{outs[0]} is assigned a value without being constrained. 
Consequently, a malicious prover could replace the value of this signal with any arbitrary value, and the verifier would still accept a proof for it. 
The correction is demonstrated in line $11$, where the operator \texttt{=} is replaced with \texttt{<==} for both assignment and constraint.
Such subtle bug could lead to significant consequences, as demonstrated by the Tornado Cash incident.

\begin{figure}[t]

\begin{lstlisting}[style=circomstyle,escapechar=@]
template MiMCSponge(nInputs, nRounds, nOutputs) {
  signal input ins[nInputs];
  signal input k;
  signal output outs[nOutputs];
  // S = R||C
  component S[nInputs + nOutputs - 1];
  for (var i = 0; i < nInputs; i++) {
    ...
  }
@\Hilightred@-  outs[0] <-- S[nInputs - 1].xL_out;
@\Hilightgreen@+  outs[0] <== S[nInputs - 1].xL_out;
  for (var i = 0; i < nOutputs - 1; i++) {
    ...
  }
}
\end{lstlisting}
\caption{Circom Circuit example: Underconstrained -- Assigned but Unconstrained}
\label{fig:assigned_but_unconstrained}
\end{figure}

\subsubsection{Underconstrained -- Missing Input Constraint}

\textbf{DSL:} Halo2 \\
\textbf{Project:} Scroll zkEVM \\
\textbf{Source:} \url{https://github.com/Zellic/publications/blob/master/Scroll%20zkEVM%20-%20Part%201%20-%20Audit%20Report.pdf} \\
\textbf{Fix:} \url{https://github.com/scroll-tech/zkevm-circuits/commit/d0e7a07e8af25220623564ef1c3ed101ce63220e}

\point{Halo2 background}
Halo2 utilizes PLONKish arithmetization, a derivative of the arithmetization introduced by PLONK~\cite{gabizon2019plonk}. 
The PLONKish arithmetization in Halo2 is coined ``UltraPLONK'' and introduces custom gates as well as lookup tables, based on plookup~\cite{gabizon2020plookup}.
In PLONKish circuits, computations are represented as tables filled with values from a finite field, where each entry, known as a cell, contributes to the circuit's overall functionality.
These circuits consist of gates, which impose constraints on specific groups of cells, thereby dictating the circuit's logic.
The table columns in Halo2 circuits are categorized into three main types: instance, advice, and fixed.
Instance columns hold public inputs, advice columns contain the private witness (comprising private inputs and intermediate values utilized in circuit computations), and fixed columns are for constant values defining the circuit.
The circuit's behavior is handled by gate constraints (polynomial constraints in the circuit), copy constraints (ensuring certain cells across the table have matching values), and lookup table constraints, all of which must be satisfied for the circuit to be valid.
Halo2, as a framework based on Plonk, uses selectors, which are binary variables that activate or deactivate gate constraints on a per-row basis.
Halo2 circuits typically include two functions called \texttt{configure} and \texttt{assign} (or \texttt{synthesize}) for setting up custom gates (i.e., constraints definition) and populating the table with values (i.e., witness generation), respectively.
Overall, Halo2 provides high composability, enabling the efficient construction and verification of complex cryptographic proofs.

A prevalent issue across various \DSLs and \eDSLs for \SNARKs is the failure to properly constrain certain inputs within a circuit. 
An illustrative case of this vulnerability is observed in the \texttt{configure} function of \texttt{LtChip} within the Scroll zkEVM, a component designed to evaluate and enforce a ``less than'' operation between two field elements (\texttt{lhs} and \texttt{rhs}). 
In Figure~\ref{fig:missing_input_constraints}, \texttt{diff} represents the byte-wise difference between \texttt{lhs} and \texttt{rhs}. 
However, the initial implementation of this function neglected to ensure that the diff values fell within the valid byte range, mistakenly leaving this responsibility to circuits incorporating the chip. 
This oversight led to significant vulnerabilities elsewhere in Scroll's codebase, where developers assumed \texttt{diff} was already constrained within the appropriate range.
The resolution involved integrating a range check within the \texttt{LtChip} to explicitly constrain \texttt{diff} to the byte range, rectifying the assumption of implicit constraint enforcement~\cite{scrollFixLtChip}. 
Such oversight is not unique to Halo2 and can also occur in other \DSLs like Circom, particularly when crucial input or output constraints are absent. 

\begin{figure}[t]
\begin{lstlisting}[style=ruststyle,escapechar=@]
pub fn configure(
    meta: &mut ConstraintSystem<F>,
    q_enable: ..., lhs: ...
    rhs: ..., u8_table: TableColumn,
) -> LtConfig<F, N_BYTES> {
    let lt = meta.advice_column();
    let diff = [(); N_BYTES].map(|_| meta.advice_column());
    let range = pow_of_two(N_BYTES * 8);
    meta.create_gate("lt gate", |meta| {
        let q_enable = q_enable(meta);
        let q_enable = q_enable.clone()(meta);
        let lt = meta.query_advice(lt, Rotation::cur());
        // get diff_bytes 
        let diff_bytes = ...
        // Check the correctness of diff_bytes
        let check_a = ...
        let check_b = bool_check(lt);
        [chec.into_iter()
            .map(move |poly| q_enable.clone() * poly)
    });
@\Hilightgreen@ +  for cell_column in diff {
@\Hilightgreen@ +      meta.lookup("range check for u8", |meta| {
@\Hilightgreen@ +        ...
@\Hilightgreen@ +      });
@\Hilightgreen@ +  }
    LtConfig {lt, diff, u8_table, range}
}
\end{lstlisting}
\caption{Halo2 Circuit example: Underconstrained -- Missing Input Constraint}
\label{fig:missing_input_constraints}
\end{figure}

\subsubsection{Underconstrained -- Unsafe Reuse of Circuit}

\textbf{DSL:} Circom \\
\textbf{Project:} circom-pairing \\
\textbf{Source:} \url{https://medium.com/veridise/circom-pairing-a-million-dollar-zk-bug-\%caught-early-c5624b278f25} \\
\textbf{Fix:} \url{https://github.com/yi-sun/circom-pairing/pull/21/commits/c686f0011f8d18e0c11bd87e0a109e9478eb9e61}

\begin{figure}[t]
\begin{lstlisting}[style=circomstyle,escapechar=@]
template CoreVerifyPubkeyG1(n, k){
    signal input pubkey[2][k];
    signal input signature[2][2][k];
    signal input hash[2][2][k];

    var q[50] = get_BLS12_381_prime(n, k);
    component lt[10];
    for(var i=0; i<10; i++){
        lt[i] = BigLessThan(n, k);
        for(var idx=0; idx<k; idx++)
            lt[i].b[idx] <== q[idx];
    }
    for(var idx=0; idx<k; idx++){
        // Assign and constraint lt[idx].a
        ...
    }
@\Hilightgreen@ +  var r = 0;
@\Hilightgreen@ +  for(var i=0; i<10; i++){
@\Hilightgreen@ +      r += lt[i].out;
@\Hilightgreen@ +  }
@\Hilightgreen@ +  r === 10;
    ...
}
\end{lstlisting}
\caption{Circom Circuit example: Underconstrained -- Unsafe Reuse of Circuit}
\label{fig:unsafe_reuse}
\end{figure}

A common root cause of vulnerabilities in circuits is the unsafe reuse of other (safe) circuits.
In Figure~\ref{fig:unsafe_reuse}, the vulnerability is related to the unsafe reuse of a sub-circuit within the \textit{circom-pairing} library in the CoreVerifyPubkeyG1 template. 
This template plays a pivotal role in verifying signatures for public keys, relying on the assumption that all inputs, specifically curve coordinates, fall within the finite field defined by the curve's prime number $q$, thus within the interval from $0$ to $q$ and are properly formatted big integers. 
The validation for this assumption is attempted through ten instances of the \texttt{BigLessThan} component, which sets its output signal \texttt{out} to $1$ if integer \texttt{a} is less than \texttt{b}, and to $0$ otherwise, also ensuring \texttt{a} and \texttt{b}'s proper format. 
However, a critical oversight was the failure to constrain the output signals of each \texttt{BigLessThan} component in the \texttt{lt} array.
Consequently, the \texttt{CoreVerifyPubkeyG1} template could accept inputs exceeding $q$ or improperly formatted, significantly enlarging the attack surface. 
The rectification for this flaw involves enforcing every \texttt{lt} component's output signal to equal \textit{one}.
This scenario underscores the inherent risk in implicitly assuming sub-circuit constraints without explicit enforcement, a common pitfall across \DSLs for \SNARKs. 
Addressing such vulnerabilities necessitates meticulous constraint application, even when optimizing, to safeguard circuit integrity. 
The example provided serves as a cautionary tale, emphasizing the importance of comprehensive constraint verification in circuit design.
Note that this problem is not unique to \SNARKs, and it is important to clarify that the \texttt{BigLessThan} circuit itself is not faulty in this context. These errors often stem from the intricacies and challenges associated with the low-level programming model of many \SNARK \DSLs. For instance, a language that warns users about unused return values could help prevent such oversights.
Similarly, another prevalent issue within the same root cause of vulnerabilities occurs when circuits explicitly outline certain assumptions -- for instance, specifying that an input should fall within a particular range -- yet these constraints are not enforced by the calling function.

\subsubsection{Underconstrained -- Out-of-Circuit Computation not Being Constrained}

\textbf{DSL:} Arkworks \\
\textbf{Project:} Penumbra \\
\textbf{Source:} \url{https://penumbra.zone/pdfs/zksecurity_penumbra_2023.pdf} \\
\textbf{Fix:} \url{https://github.com/penumbra-zone/penumbra/commit/e019839939968012ed2d24cf65bdd86d239b50e9}

\begin{figure}[t]
\begin{lstlisting}[style=ruststyle]
pub fn derive(nk: &NullifierKeyVar, ak: &AuthorizationKeyVar) -> Result<Self, SynthesisError> {
    let cs = nk.inner.cs();
    let ivk_domain_sep = FqVar::new_constant(cs.clone(), *IVK_DOMAIN_SEP)?;
    let ivk_mod_q = poseidon377::r1cs::hash_2(
        cs.clone(),
        cs,
        &ivk_domain_sep,
        (nk.inner.clone(), ak.inner.compress_to_field()?),
    )?;
    let inner_ivk_mod_q: Fq = ivk_mod_q.value().unwrap_or_default();
    let ivk_mod_r = Fr::from_le_bytes_mod_order(&inner_ivk_mod_q.to_bytes());
    let ivk = NonNativeFieldVar::<Fr, Fq>::new_variable(
        cs,
        || Ok(ivk_mod_r),
        AllocationMode::Witness,
    )?;
    Ok(IncomingViewingKeyVar { inner: ivk })
}
\end{lstlisting}
\caption{Arkworks Circuit example: Underconstrained -- Out-of-Circuit Computation not Being Constrained}
\label{fig:out-of-circuit}
\end{figure}

Penumbra uses nullifiers in shielded transactions to ensure notes can be spent without revealing their spend status. 
A nullifier is deterministically derived using a nullifier key (\texttt{nk}) and a note commitment. 
To spend a note, users must prove ownership of an incoming viewing key (\texttt{ivk}), which is derived from \texttt{nk}. 
The \texttt{ivk} derivation involves converting a value from the circuit field ($F_q$) to a non-native field ($F_r$), as \texttt{ivk} is used for scalar multiplication (see Figure~\ref{fig:out-of-circuit}).
The vulnerability arises during this field conversion, where the \texttt{ivk\_mod\_q} value is taken out of the circuit, converted to \texttt{ivk\_mod\_r}, and then reinserted as a witness value under \texttt{ivk}, removing any constraint linking \texttt{ivk} back to \texttt{ivk\_mod\_q}. 
This lapse allows a malicious prover to manipulate \texttt{ivk} or \texttt{ivk\_mod\_q}, enabling the creation of incorrect nullifiers and potential double-spending.

%

\subsection{Integration Layer Vulnerability Examples}
\label{app:integration}

\subsubsection{Passing Unchecked Data -- Missing Validation Input}

\begin{figure}
\begin{lstlisting}[style=soliditystyle,escapechar=@]
function collectAirdrop(bytes calldata proof, bytes32 nullifierHash) public {
@\Hilightgreen@ + require(uint256(nullifierHash) < SNARK_FIELD ,"...");
    require(!nullifierSpent[nullifierHash], "...");

    uint[] memory pubSignals = new uint[](3);
    pubSignals[0] = uint256(root);
    pubSignals[1] = uint256(nullifierHash);
    pubSignals[2] = uint256(uint160(msg.sender));
    require(verifier.verifyProof(proof, pubSignals), "...");
    nullifierSpent[nullifierHash] = true;
    airdropToken.transfer(msg.sender, amountPerRedemption);
}
\end{lstlisting}
\caption{Integration example: Passing Unchecked Data -- Missing Validation Input}
\label{fig:integration-passing}
\end{figure}

\textbf{Language} Solidity \\
\textbf{Project:} zkdrops \\
\textbf{Source:} \url{https://github.com/a16z/zkdrops/pull/2} \\
\textbf{Fix:} \url{https://github.com/a16z/zkdrops/pull/2/files}

Zkdrops is a tool designed for distributing tokens, commonly known as airdrops, in a privacy-preserving way. 
Users initially submit a \textit{commitment} message publicly and later claim their tokens through a \ZKP, verifying their inclusion in a Merkle tree without revealing their identity. 
This process blends their claim with others, maintaining anonymity. 
To prevent double-spending, zkdrops employ \textit{nullifiers}; users must submit a unique nullifier with their proof to collect tokens. 
However, a flaw was identified in the system's implementation, showcased in the provided Solidity code (c.f.\ Figure~\ref{fig:integration-passing}). 
The issue lies in the failure to validate that the nullifier falls within the SNARK field's bounds. 
This enables the creation of nullifiers that, while valid uint256 have the same result modulo the field, are treated as distinct by the spent nullifier dictionary, thereby permitting double-spending.
The fix of this issue involves incorporating a validation step (shown in line 3 of Figure~\ref{fig:integration-passing}) to ensure nullifier values are properly constrained. 
This type of vulnerability is common at the integration layer, underscoring the importance for developers to rigorously apply and check all assumptions made at the circuit level within the integration layer to prevent crucial vulnerabilities.

\subsubsection{Proof Delegation Error -- Integration Design Error}

\begin{figure}
\begin{lstlisting}[style=ruststyle,escapechar=@]
pub struct Execution<N: Network> {
    transitions: IndexMap<N::TransitionID, Transition<N>>,
    global_state_root: N::StateRoot,
    proof: Option<Proof<N>>,
}
\end{lstlisting}
\caption{Integration example: Proof Delegation Error -- Integration Design Error}
\label{fig:integration-delegation}
\end{figure}

\textbf{Language} Rust \\
\textbf{Project:} AleoVM \\
\textbf{Source:} \url{https://www.zksecurity.xyz/blog/2023-aleo-synthesizer.pdf}

To facilitate a function call on the Aleo network, a user decomposes the call into a sequence of transitions (see Figure~\ref{fig:integration-delegation}). 
Each transition correlates with a nested function call, implying that if the initial function call, referred to as \texttt{main}, generates $n$ nested calls, there would be $n+1$ transitions, with the last transition reflecting the completion of the main function call.
Transitions are linked to specific synthesized circuits, and each circuit is paired with a proof, though all proofs are ultimately merged into a single comprehensive proof. 
The process for each transition circuit begins with the verification of a signed request, confirming the user's intention. 
This mechanism enables a user to \textit{delegate} the execution of a particular function to a third-party prover without granting them the authority to execute additional functions.
Since each transition possesses an individual request, detached from the overall execution context, a third-party prover can sequentially exclude callers by omitting subsequent requests during execution. 
This scenario permits a third-party prover to truncate the full execution desired by the user, only processing a limited number of nested calls and possibly hindering the execution of encompassing logic. 
For instance, in a situation where function \texttt{fn1} invokes \texttt{fn2}, the third-party prover might only acknowledge the request to execute \texttt{fn2}.
Although this vulnerability is not explicitly tied to \ZKPs, it highlights challenges within the broader infrastructure necessary for implementing and utilizing \ZKPs effectively. 
The issue arises from users needing to entrust execution to potentially untrusted provers, which, due to flawed design, allows these provers to alter the intended execution path. 
Addressing this requires a redesign incorporating a counter in each transition circuit as a publicly verifiable input. 
This counter, incremented with each transition proof verification, must be included in each request's signature, ensuring the integrity and intended sequence of execution.

\end{document}

%% file: packages.tex
\usepackage{cite}
\usepackage{amsmath,amssymb,amsfonts}
\usepackage{algorithmic}
\usepackage{textcomp}
\usepackage{bigstrut}
\usepackage{etoolbox}
\usepackage{graphicx}
\usepackage{multirow}
\usepackage{environ}
\usepackage{xcolor}
\usepackage[acronym]{glossaries} 
\usepackage{pifont}
\usepackage{sepnum}
\usepackage{xspace}
\usepackage[group-separator={,}]{siunitx}
\usepackage[most]{tcolorbox}
\usepackage{subcaption}
\usepackage{bigstrut}
\usepackage{makecell}
\usepackage{booktabs}
\usepackage{enumitem}
\usepackage{listings}
\usepackage{amsthm}
\PassOptionsToPackage{hyphens}{url}\usepackage{hyperref}
\usepackage[T1]{fontenc}
\usepackage{wasysym}
\usepackage{color}
\usepackage{listings}
\usepackage{listings-rust}
\usepackage{courier}

\newcommand{\empirical}[1]{#1}

\usepackage{colortbl}
\definecolor{gainsboro}{rgb}{0.86, 0.86, 0.86}

\newcommand{\point}[1]{\par\smallskip\noindent\textbf{#1.}}

\newcounter{vulnerabilityID}
\newcommand{\vuln}[1]{%
  \stepcounter{vulnerabilityID}
  \smallskip\noindent\textbf{V\arabic{vulnerabilityID}. #1:}
}

\newcounter{rootID}
\newcommand{\rootcause}[1]{%
  \stepcounter{rootID}
  \smallskip\noindent\textbf{R\arabic{rootID}. #1:}
}

\theoremstyle{definition}

\newcommand{\fc}{\CIRCLE}
\newcommand{\ec}{\Circle}

\newcommand{\tk}{\ding{52}}
\newcommand{\cx}{\ding{56}}
\newcommand{\mb}{\ding{73}}
\NewDocumentCommand{\rot}{O{45} O{1em} m}{\makebox[#2][l]{\rotatebox{#1}{#3}}}

\definecolor{codegreen}{rgb}{0.56, 0.93, 0.56}
\definecolor{codered}{rgb}{0.94, 0.5, 0.5}

\definecolor{keyword1}{RGB}{75,119,190} 
\definecolor{keyword2}{RGB}{30,135,92}
\definecolor{keyword3}{RGB}{180,95,75}
\definecolor{comment}{RGB}{38,166,91} 
\definecolor{bgcolor}{RGB}{246,246,246} 
\definecolor{Red}{RGB}{255,0,0}

\definecolor{verylightgray}{rgb}{.97,.97,.97}

\lstdefinelanguage{Solidity}{
	keywords=[1]{anonymous, assembly, assert, balance, break, call, callcode, case, catch, class, constant, continue, constructor, contract, debugger, default, delegatecall, delete, do, else, emit, event, experimental, export, external, false, finally, for, function, gas, if, implements, import, in, indexed, instanceof, interface, internal, is, length, library, log0, log1, log2, log3, log4, memory, modifier, new, payable, pragma, private, protected, public, pure, push, require, return, returns, revert, selfdestruct, send, solidity, storage, struct, suicide, super, switch, then, this, throw, transfer, true, try, typeof, using, value, view, while, with, addmod, ecrecover, keccak256, mulmod, ripemd160, sha256, sha3}, 
	keywordstyle=[1]\color{blue}\bfseries,
	keywords=[2]{address, bool, byte, bytes, bytes1, bytes2, bytes3, bytes4, bytes5, bytes6, bytes7, bytes8, bytes9, bytes10, bytes11, bytes12, bytes13, bytes14, bytes15, bytes16, bytes17, bytes18, bytes19, bytes20, bytes21, bytes22, bytes23, bytes24, bytes25, bytes26, bytes27, bytes28, bytes29, bytes30, bytes31, bytes32, enum, int, int8, int16, int24, int32, int40, int48, int56, int64, int72, int80, int88, int96, int104, int112, int120, int128, int136, int144, int152, int160, int168, int176, int184, int192, int200, int208, int216, int224, int232, int240, int248, int256, mapping, string, uint, uint8, uint16, uint24, uint32, uint40, uint48, uint56, uint64, uint72, uint80, uint88, uint96, uint104, uint112, uint120, uint128, uint136, uint144, uint152, uint160, uint168, uint176, uint184, uint192, uint200, uint208, uint216, uint224, uint232, uint240, uint248, uint256, var, void, ether, finney, szabo, wei, days, hours, minutes, seconds, weeks, years},	
	keywordstyle=[2]\color{teal}\bfseries,
	keywords=[3]{block, blockhash, coinbase, difficulty, gaslimit, number, timestamp, msg, data, gas, sender, sig, value, now, tx, gasprice, origin},	
	keywordstyle=[3]\color{violet}\bfseries,
	identifierstyle=\color{black},
	sensitive=true,
	comment=[l]{//},
	morecomment=[s]{/*}{*/},
	commentstyle=\color{gray}\ttfamily,
	stringstyle=\color{red}\ttfamily,
	morestring=[b]',
	morestring=[b]"
}

\lstdefinestyle{examplestyle}{
    basicstyle=\footnotesize\ttfamily,
    breakatwhitespace=false,
    breaklines=true,
    captionpos=b,
    keepspaces=true,
    numbers=left,
    numbersep=5pt,
    showspaces=false,
    showstringspaces=false,
    showtabs=false,
    tabsize=2,
    language={},
    moredelim=**[is][\color{codered}]{@}{@},
    moredelim=**[is][\color{codegreen}]{$}{$},
}

\lstdefinestyle{ruststyle}{
    basicstyle=\footnotesize\ttfamily,
    breakatwhitespace=false,
    breaklines=true,
    captionpos=b,
    keepspaces=true,
    numbers=left,
    numbersep=5pt,
	numberstyle=\tiny\bfseries,
    showspaces=false,
    showstringspaces=false,
	showlines=true,
    showtabs=false,
    tabsize=2,
	xleftmargin=0pt,
    xrightmargin=0pt,
    columns=fullflexible,
    backgroundcolor=\color{bgcolor},
    language=Rust,
}

\lstdefinestyle{soliditystyle}{
    basicstyle=\footnotesize\ttfamily,
    breakatwhitespace=false,
    breaklines=true,
    captionpos=b,
    keepspaces=true,
    numbers=left,
    numbersep=5pt,
	numberstyle=\tiny\bfseries,
    showspaces=false,
    showstringspaces=false,
	showlines=true,
    showtabs=false,
    tabsize=2,
	xleftmargin=0pt,
    xrightmargin=0pt,
    columns=fullflexible,
    backgroundcolor=\color{bgcolor},
    language=Solidity,
}

\lstdefinestyle{circomstyle}{
    alsoletter={?!-*<=>:+/.},
    morekeywords=[1]{template,signal,input,output,var,for,component,pragma,circom,include,public},
    morekeywords=[2]{==,?,:,+,*,-,=,<,++,/},
    morekeywords=[3]{==>,<--,===,<==,main},
    backgroundcolor=\color{bgcolor},
	numbers=left,
	numberstyle=\tiny\bfseries,
	numbersep=5pt,
	showstringspaces=false,
	tabsize=2,
	basicstyle=\footnotesize\ttfamily,
	commentstyle=\color{comment}\textit,
	morecomment=[l]{//},
	keywordstyle=[1]{\color{keyword1}\bfseries},
	keywordstyle=[2]{\color{keyword2}\bfseries},
	keywordstyle=[3]{\color{keyword3}\bfseries},
	showlines=true,
	xleftmargin=0pt,
    xrightmargin=0pt,
    columns=fullflexible,
	mathescape
}

%% file: acronyms.tex
\newacronym{W3C}{W3C}{World-Wide-Web Consortium}
\newacronym{SNARK}{SNARK}{Succinct Non-interactive Arguments of Knowledge}
\newacronym{SNARKs}{SNARKs}{Succinct Non-interactive Arguments of Knowledge}
\newacronym{ZKPs}{ZKPs}{Zero-Knowledge Proofs}
\newacronym{IOP}{IOP}{Interactive Oracle Proof}
\newacronym{STARKs}{STARKs}{Succinct Transparent Arguments of Knowledge}
\newacronym{EVM}{EVM}{Ethereum Virtual Machine}
\newacronym{FPUs}{FPUs}{Floating Point Units}
\newacronym{FFT}{FFT}{Fast Fourier Transform}
\newacronym{DLP}{DLP}{Discrete Logarithm Problem}
\newacronym{R1CS}{R1CS}{Rank-1 Constraint System}
\newacronym{KZG}{KZG}{Kate, Zaverucha, Goldberg}
\newacronym{PSE}{PSE}{Privacy \& Scaling Explorations}
\newacronym{DSL}{DSL}{Domain Specific Language}
\newacronym{DSLs}{DSLs}{Domain Specific Languages}
\newacronym{AIR}{AIR}{Arithmetic Intermediate Representation}
\newacronym{MPC}{MPC}{Multi Party Computation}
\newacronym{FHE}{FHE}{Fully Homomorphic Encryption}

%% file: macros.tex
\newtoggle{doPrintCommentsAll}
\NewEnviron{maybePrintAll}{%
  \iftoggle{doPrintCommentsAll}{\BODY}{}%
}

\toggletrue{doPrintCommentsAll}

\definecolor{applegreen}{rgb}{0.55, 0.71, 0.0}

\newcommand{\extended}[1]{#1}

\newcommand{\ZKP}{ZKP\xspace}
\newcommand{\ZKPs}{\gls{ZKPs}\xspace}
\newcommand{\ZK}{ZK\xspace}
\newcommand{\SNARKs}{\gls{SNARKs}\xspace}
\newcommand{\SNARK}{SNARK\xspace}

\newcommand{\DSL}{DSL\xspace}
\newcommand{\DSLs}{\gls{DSLs}\xspace}

\newcommand{\eDSLs}{eDSLs\xspace}

\newcommand{\totalvulns}{$141$\xspace}
\newcommand{\totalaudits}{$107$\xspace}
\newcommand{\totalzkaudits}{$75$\xspace}
\newcommand{\circuitsoundness}{$95$\xspace}
\newcommand{\circuitcompleteness}{$4$\xspace}

\newcommand{\CircuitLayer}{Circuit Layer\xspace}
\newcommand{\FrontendLayer}{Frontend Layer\xspace}
\newcommand{\BackendLayer}{Backend Layer\xspace}
\newcommand{\IntegrationLayer}{Integration Layer\xspace}

%% file: figures/codeListing.tex








\definecolor{gainsboro}{rgb}{0.86, 0.86, 0.86}

\begin{figure}[!t]
  \centering
  \scriptsize
  \fbox{
  \begin{minipage}{\columnwidth}
  \linespread{1.2}\selectfont
  \texttt{type \textbf{Circuit} struct \{ } \newline
  \hspace*{1em} \texttt{X frontend.Variable \textcolor{blue}{\textasciigrave gnark:",private"\textasciigrave}} \newline
  \hspace*{1em} \texttt{C frontend.Variable \textcolor{blue}{\textasciigrave gnark:",public"\textasciigrave}} \newline
  \hspace*{1em} \texttt{Y frontend.Variable \textcolor{blue}{\textasciigrave gnark:",public"\textasciigrave}} \newline
  \texttt{\}} \newline
  \texttt{func (circuit *Circuit) Define(api frontend.API) \textbf{error} \{} \newline
  \hspace*{0.5em} \texttt{outputs := api.Compiler().NewHint(hint.SqrtHint, 1, circuit.X)} \newline
  \hspace*{0.5em} \texttt{squareRoot := outputs[0]} \newline
  \hspace*{0.5em} \colorbox{gainsboro}{\texttt{api.AssertIsEqual(api.Mul(squareRoot, squareRoot), circuit.X)}} \newline
  \hspace*{0.5em} \colorbox{gainsboro}{\texttt{result := api.Mul(squareRoot, circuit.Const)}} \newline
  \hspace*{0.5em} \colorbox{gainsboro}{\texttt{api.AssertIsEqual(circuit.Y, result)}} \newline
  \hspace*{0.5em} \texttt{return nil} \newline
  \texttt{\}}
  \end{minipage}
  }
  \caption{
  Example circuit written in the eDSL gnark~\cite{gnark}. Lines highlighted in gray add constraints in the circuit.
  }
  \label{fig:golang_code}
\end{figure}

%% file: tables/adversarialKnowledge.tex
\begin{table}[!t]
\centering
\resizebox{\columnwidth}{!}{ 
\begin{tabular}{@{}l c@{\hspace{1.5em}}c@{\hspace{1.5em}}c@{\hspace{1.5em}}c@{\hspace{1.5em}}c@{\hspace{1.5em}}c@{\hspace{1.5em}}c@{\hspace{1.5em}}c@{\hspace{1.5em}}c@{\hspace{1.5em}} c@{\hspace{1.5em}}}
\toprule
  \textbf{Adversarial Role} & 
  \rot[25]{Public Input} & 
  \rot[25]{Private Input} & 
  \rot[25]{Circuit} & 
  \rot[25]{Public Witness} & 
  \rot[25]{Private Witness} & 
  \rot[25]{Arithmetization} & 
  \rot[25]{CRS} & 
  \rot[25]{Prover Key} & 
  \rot[25]{Verifier Key} & 
  \rot[25]{Proof} \\
\midrule
$R_1$ - Network Adversary & \mb & \cx & \tk & \tk & \cx & \mb & \tk & \mb & \mb & \tk \\
$R_2$ - Adversarial User  & \tk & \tk & \tk & \tk & \tk & \mb & \tk & \mb & \mb & \mb \\
$R_3$ - Adversarial Prover  & \tk & \mb & \tk & \tk & \mb & \tk & \tk & \tk & \mb & \tk\\
\bottomrule
\end{tabular}
}
\caption{
Categorization of adversarial roles by the knowledge they can obtain and utilize. A \textit{network} adversary can observe existing proofs and reuse them with different inputs to exploit malleability vulnerabilities. The \textit{user} can delegate proof generation to a proving service, while the \textit{prover} can generate and has complete control over proof generation. Typically, soundness vulnerabilities due to circuit bugs can exploited only by the prover. Adversary has ``\tk'' knowledge, ``\cx'' no knowledge, ``\mb'' maybe knowledge.}
\label{tab:adversarial_knowledge}
\end{table}

%% file: tables/sources_table.tex
\begin{table}[t]
\centering
\footnotesize
\setlength{\tabcolsep}{3pt}
\begin{tabular}{lrrrr}
\toprule
\multirow{2}{*}{\bf Layer} & \bf Security & \bf Vulnerability & \bf Bug & \multirow{2}{*}{\bf Total} \bigstrut[t] \\ 
& \bf audits & \bf disclosures & \bf Tracker & \\ 
\midrule
Integration & 8 & 1 & 4 & 13 \bigstrut[t] \\
Circuit & 86 & 10 & 3 & 99 \\
Frontend & 0 & 0 & 6 & 6 \\
Backend & 7 & 5 & 11 & 23 \\
\midrule
\bf Total & \bf 101 & \bf 16 & \bf 24 & \bf 141 \bigstrut[t] \\
\bottomrule
\end{tabular}
\caption{Origins of vulnerability reports.}
\label{tab:sources}
\end{table}

%% file: tables/impact_table.tex
\begin{table}[t]
\centering
\footnotesize
\setlength{\tabcolsep}{3pt}
\begin{tabular}{lrrr}
\toprule
\textbf{Impact} & \textbf{Soundness} & \textbf{Completeness} & \textbf{Zero Knowledge} \\
\midrule
Integration & 11 & 2 & 0 \\
Circuit & 94 & 5 & 0 \\
Frontend & 2 & 4 & 0 \\
Backend & 17 & 3 & 3 \\
\midrule
\bf Total & \bf 124 & \bf 14 & \bf 3 \\
\bottomrule
\end{tabular}
\caption{Impact of SNARK vulnerabilities.}
\label{tab:impact}
\end{table}

%% file: tables/circuits.tex
\begin{table}[tb]
\centering
\footnotesize
\resizebox{1\linewidth}{!}{
\begin{tabular}{lrrrr}
\toprule
\bf Root Cause & \bf UC & \bf OC & \bf CE & \bf Total \bigstrut\\
\midrule
Assigned but Unconstrained & 14 & 0 & 0 & 14 \\
Missing Input Constraints & 25 & 0 & 0 & 25 \\
Unsafe Reuse of Circuit & 9 & 0 & 0 & 9 \\
Wrong translation of logic into constraints & 32 & 0 & 2 & 34 \\
Incorrect Custom Gates & 1 & 0 & 0 & 1 \\
Out-of-Circuit Computation Not Being Constrained & 1 & 0 & 0 & 1 \\
Arithmetic Field Errors & 8 & 0 & 0 & 8 \\
Bad Circuit/Protocol Design & 4 & 0 & 0 & 4 \\
Other Programming Errors & 1 & 1 & 1 & 3 \\
\midrule
\bf Total & \bf 95 & \bf 1 & \bf 3 & \bf 99 \\
\bottomrule
\end{tabular}
}
\caption{Circuit vulnerabilities. UC: Under-Constrained, OC: Over-Constrained, CI, Computational/Hints Error.}
\label{tab:circuits}
\end{table}

%% file: tables/application.tex
\begin{table}[t]
\centering
\footnotesize
\resizebox{1\linewidth}{!}{
\begin{tabular}{lccccr}
\hline
\textbf{Root Cause} & \textbf{PDE} & \textbf{PCE} & \textbf{PUD} & \textbf{ZKPCLE} & \textbf{Total} \bigstrut\\
\hline
Integration Design Error & 1 & 2 & 0 & 0 & 3 \bigstrut[t]\\
Missing Validation Input & 0 & 0 & 9 & 1 & 10 \\
\hline
\textbf{Total} & \textbf{1} & \textbf{2} & \textbf{9} & \textbf{1} & \textbf{13} \bigstrut\\
\hline
\end{tabular}
}
\caption{Integration vulnerabilities. PDE: Proof Delegation Error, PCE: Proof Composition Error, PUD: Passing Unchecked Data, ZKPCLE: ZKP Complementary Logic Error}
\label{tab:applications}
\end{table}

%% file: tables/framework.tex
\begin{table}[t]
\centering
\footnotesize
\setlength{\tabcolsep}{7pt}
\begin{tabular}{llr}
\toprule
\bf Layer & \bf Vulnerability & \bf \# Bugs \\
\midrule
\multirow{3}{*}{Frontend} & Incorrect Constraint Compilation & 3 \\
 & Witness Generation Error & 3 \\
 \midrule
 & \bf Total & \bf 6 \\
\midrule
\multirow{4}{*}{Backend} & Setup Error & 1 \\
 & Prover Error & 6 \\
 & Unsafe Verifier & 16 \\
\midrule
 & \bf Total & \bf 23 \\
\bottomrule
\end{tabular}
\caption{Summary of frontend and backend bugs.}
\label{tab:framework}
\end{table}

%% file: tables/defenses.tex
\begin{table*}[t]
\centering
\footnotesize
\resizebox{1\linewidth}{!}{
\begin{tabular}{|ll|ccccccccccc|}
\hline
 & \bf Vulnerability & \multicolumn{11}{c|}{\bf Defenses} \bigstrut[t]\\
 & 
 & \rotatebox{65}{Circomspect~\cite{circomspect}} 
 & \rotatebox{65}{ZKAP~\cite{wen2023practical}} 
 & \rotatebox{65}{Korrekt~\cite{soureshjani2023automated}} 
 & \rotatebox{65}{Coda~\cite{liu2023certifying}} 
 & \rotatebox{65}{Ecne~\cite{wang2022ecne}} 
 & \rotatebox{65}{Picus~\cite{pailoor2023automated}} 
 & \rotatebox{65}{Aleo~\cite{coglio2023compositional,coglio2023formal}} 
 & \rotatebox{65}{OWBB23~\cite{ozdemir2023bounded}}
 & \rotatebox{65}{SnarkProbe~\cite{fansnarkprobe}}
 & \rotatebox{65}{CIVER~\cite{isabel2024scalable}}
 & \rotatebox{65}{Trad Sec*} \\
\hline\hline
\multirow{3}{*}{\rotatebox[origin=c]{90}{\parbox[c]{1cm}{\centering \bf Circuit}}} 
 & Under-Constrained & \fc & \fc & \fc & \fc & \fc & \fc & \fc & \ec & \fc & \fc & \ec \bigstrut[t]\\
 & Over-Constrained & \ec & \ec & \fc & \fc & \ec & \ec & \fc & \ec & \fc & \ec & \ec \\
 & Computational Error & \ec & \ec & \ec & \fc & \ec & \ec & \fc & \ec & \ec & \ec & \fc \\
\hline
\multirow{2}{*}{\rotatebox[origin=c]{90}{\parbox[c]{1.3cm}{\centering \bf Frontend}}} 
 & \multirow{2}{*}{Incorrect Constraint Compilation} & \multirow{2}{*}{\ec} & \multirow{2}{*}{\ec} & \multirow{2}{*}{\ec} & \multirow{2}{*}{\ec} & \multirow{2}{*}{\ec} & \multirow{2}{*}{\ec} & \multirow{2}{*}{\ec} & \multirow{2}{*}{\fc} & \multirow{2}{*}{\ec} & \multirow{2}{*}{\ec} & \multirow{2}{*}{\ec} \\
 & & & & & & & & & & & & \\
 & \multirow{2}{*}{Witness Generation Error} & \multirow{2}{*}{\ec} & \multirow{2}{*}{\ec} & \multirow{2}{*}{\ec} & \multirow{2}{*}{\ec} & \multirow{2}{*}{\ec} & \multirow{2}{*}{\ec} & \multirow{2}{*}{\ec} & \multirow{2}{*}{\ec} & \multirow{2}{*}{\ec} & \multirow{2}{*}{\ec} & \multirow{2}{*}{\fc} \\
 & & & & & & & & & & & & \\
\hline
\multirow{3}{*}{\rotatebox[origin=c]{90}{\parbox[c]{1.2cm}{\centering \bf Backend}}} 
 & Setup Error & \ec & \ec & \ec & \ec & \ec & \ec & \ec & \ec & \fc & \ec & \ec \\
 & Prover Error & \ec & \ec & \ec & \ec & \ec & \ec & \ec & \ec & \ec & \ec & \fc \\
 & Unsafe Verifier & \ec & \ec & \ec & \ec & \ec & \ec & \ec & \ec & \ec & \ec & \fc  \bigstrut \\
\hline
\end{tabular}
}
\caption{An overview of papers and tools offering defense mechanisms for addressing vulnerabilities in SNARK-based systems. While numerous techniques are concentrated on circuit-layer security, they often face challenges in scalability or possess limited functionality tied to specific DSLs. 
The last column represents traditional security tools (e.g., AFL and property-based testing) employed by auditors in audit reports. Note that these tools typically have limited capabilities for detecting SNARK-related bugs. We excluded the integration layer from the table as no defense detects bugs in that layer.}
\label{tab:snark-defenses}
\end{table*}

%% file: tables/examples_table.tex
\begin{table*}[t]
\centering
\begin{tabular}{ccp{9cm}cc}
\hline
\bf Ref & \bf Year & \bf Description & \bf Layer & \bf Impact  \bigstrut\\
\hline
\cite{zcash2019counterfeiting} & 2019 & The trusted setup in Zcash's Pinocchio-based implementation produced additional parameters that compromised the 'toxic waste' secrecy. This vulnerability enabled malicious provers to create false proofs, leading to the unauthorized creation of Zcash tokens, effectively resulting in token counterfeiting. & Proof System & Breaking Soundness \bigstrut[t] \\
~\cite{gurkan2019tornadocash} & 2019 & A vulnerability in circomlib's implementation of the MiMC hash function, used in Tornado's merkle tree of deposits, permitted the verification of incorrect hashes. This flaw could lead to unauthorized withdrawals of 0.1 ETH from Tornado.cash without legitimate deposits. The core issue was a computation in the hash function being performed without the necessary constraint in place, enabling this exploit. & Circuit & Breaking Soundness \\
\cite{quan2021plonkcpp} & 2021 & The Aztec Plonk verifier had a flaw where setting two key elements to 0 caused it to mistakenly accept a proof, as it misinterpreted these zeros as the `point at infinity' in elliptic curve cryptography. This vulnerability enables attackers to forge proofs by exploiting the verifier's handling of these specific zero values.
 & Backend & Breaking Soundness \\
~\cite{charbonnet2022semaphore} & 2022 & In Solidity, the uint256 type's capacity to exceed the snark scalar field order poses an overflow risk. To counter this, the Semaphore verifier smart contract fails if a public input, like a group ID, exceeds this order. If a group's ID is set too high, it causes the verifier to revert, making the group non-functional. Hence, the Semaphore smart contract had to perform a check before passing data to the circuits. & Integration & Breaking Completeness \\
\hline
\end{tabular}
\caption{Representative vulnerabilities, exploits, and whitehat attacks for SNARKs, showing the diversity of affected components and their security impact.}
\label{tab:examples}
\end{table*}